\documentclass[aps,prd,reprint,superscriptaddress,showpacs,showkeys]{revtex4-1}

\usepackage{graphicx}
\usepackage{natbib}
\usepackage{color}

\usepackage{subfigure}
\usepackage{array}
\usepackage{url}
\usepackage{hyperref}

\usepackage{multirow} 

\usepackage{enumitem}
\setenumerate{itemindent=2em,leftmargin=0pt,itemsep=0pt,partopsep=0pt}
\setitemize[1]{itemindent=2em,leftmargin=0pt,itemsep=2pt,partopsep=0pt,topsep=0pt}

\begin{document}


\title{Study on cosmogenic activation in germanium detectors for future tonne-scale CDEX experiment}


\affiliation{Key Laboratory of Particle and Radiation Imaging (Ministry of Education) and Department of Engineering Physics, Tsinghua University, Beijing 100084}
\affiliation{Department of Physics, Tsinghua University, Beijing 100084}
\affiliation{College of Physical Science and Technology, Sichuan University, Chengdu 610064}
\affiliation{Institute of Physics, Academia Sinica, Taipei 11529}
\affiliation{Department of  Physics, Beijing Normal University, Beijing 100875}
\affiliation{Center for High Energy Physics, Tsinghua University, Beijing 100084}
\affiliation{Collaborative Innovation Center of Quantum Matter, Beijing 100084}

\author{J.~L.~Ma}
\affiliation{Key Laboratory of Particle and Radiation Imaging (Ministry of Education) and Department of Engineering Physics, Tsinghua University, Beijing 100084}
\affiliation{Department of Physics, Tsinghua University, Beijing 100084}
\author{Q.~Yue}
\email{Corresponding author: yueq@mail.tsinghua.edu.cn}
\affiliation{Key Laboratory of Particle and Radiation Imaging (Ministry of Education) and Department of Engineering Physics, Tsinghua University, Beijing 100084}
\author{S.~T.~Lin}
\affiliation{College of Physical Science and Technology, Sichuan University, Chengdu 610064}
\author{H.~T.~Wong}
\affiliation{Institute of Physics, Academia Sinica, Taipei 11529}
\author{J.~W.~Hu}
\affiliation{Key Laboratory of Particle and Radiation Imaging (Ministry of Education) and Department of Engineering Physics, Tsinghua University, Beijing 100084}
\author{L.~P.~Jia}
\affiliation{Key Laboratory of Particle and Radiation Imaging (Ministry of Education) and Department of Engineering Physics, Tsinghua University, Beijing 100084}
\author{H.~Jiang}
\affiliation{Key Laboratory of Particle and Radiation Imaging (Ministry of Education) and Department of Engineering Physics, Tsinghua University, Beijing 100084}
\author{J.~Li}
\affiliation{Key Laboratory of Particle and Radiation Imaging (Ministry of Education) and Department of Engineering Physics, Tsinghua University, Beijing 100084}
\author{S.~K.~Liu}
\affiliation{College of Physical Science and Technology, Sichuan University, Chengdu 610064}
\author{Z.~Z.~Liu}
\affiliation{Key Laboratory of Particle and Radiation Imaging (Ministry of Education) and Department of Engineering Physics, Tsinghua University, Beijing 100084}
\author{H.~Ma}
\email{Corresponding author: mahao@mail.tsinghua.edu.cn}
\affiliation{Key Laboratory of Particle and Radiation Imaging (Ministry of Education) and Department of Engineering Physics, Tsinghua University, Beijing 100084}
\author{W.~Y.~Tang}
\affiliation{Key Laboratory of Particle and Radiation Imaging (Ministry of Education) and Department of Engineering Physics, Tsinghua University, Beijing 100084}
\author{Y.~Tian}
\affiliation{Key Laboratory of Particle and Radiation Imaging (Ministry of Education) and Department of Engineering Physics, Tsinghua University, Beijing 100084}
\author{L.~Wang}
\affiliation{Department of  Physics, Beijing Normal University, Beijing 100875}
\author{Q.~Wang}
\affiliation{Key Laboratory of Particle and Radiation Imaging (Ministry of Education) and Department of Engineering Physics, Tsinghua University, Beijing 100084}
\affiliation{Center for High Energy Physics, Tsinghua University, Beijing 100084}
\affiliation{Collaborative Innovation Center of Quantum Matter, Beijing 100084}
\author{L.~T.~Yang}
\affiliation{Key Laboratory of Particle and Radiation Imaging (Ministry of Education) and Department of Engineering Physics, Tsinghua University, Beijing 100084}
\affiliation{Department of Physics, Tsinghua University, Beijing 100084}
\author{Z.~Zeng}
\affiliation{Key Laboratory of Particle and Radiation Imaging (Ministry of Education) and Department of Engineering Physics, Tsinghua University, Beijing 100084}
\noaffiliation



\date{August 18, 2018}

\begin{abstract}
A study on cosmogenic activation in germanium was carried out to evaluate the cosmogenic background level of natural and $^{70}$Ge depleted germanium detectors. The production rates of long-lived radionuclides were calculated with Geant4 and CRY. Results were validated by comparing the simulated and experimental spectra of CDEX-1B detector. Based on the validated codes, the cosmogenic background level was predicted for further tonne-scale CDEX experiment. The suppression of cosmogenic background level could be achieved by underground germanium crystal growth and high-purity germanium detector fabrication to reach the sensitivity requirement for direct detection of dark matter. With the low cosmogenic background, new physics channels, such as solar neutrino research and neutrinoless double-beta decay experiments, were opened and the corresponding simulations and evaluations were carried out.
\end{abstract}


\maketitle

\section{Introduction}\label{section1}

For rare event experiments, such as direct detection of Weakly Interacting Massive Particles (WIMPs) and search for neutrinoless double-beta ($0\nu\beta\beta$) decay, background level has always been the focus of attention and the key point of experiments. High-purity Germanium (HPGe) is a good target and detector that can be used in both experiments. Since a lot of efforts have been devoted to suppress the background of the HPGe detector system, including operating in deep underground laboratories, carefully selecting low radioactive materials, producing necessary parts underground, and building active and passive shields, the background level can be typically decreased to about 1 count/(keV$\cdot$kg$\cdot$day) (cpkkd) for direct light dark matter detection at 2 $\sim$ 4 keV energy range \cite{ref1,ref2} and about 10$^{-6}$ cpkkd for $0\nu\beta\beta$ decay detection at Q$_{\beta\beta}$ of 2039 keV in recent years \cite{ref3,ref4,ref5,ref6}. In addition to primordial radionuclides, cosmogenic radionuclides in crystals can contribute crucial backgrounds. Reliable calculations are needed to evaluate cosmogenic background and provide projections on future tonne-scale HPGe experiments.

The China Dark matter EXperiment (CDEX) Collaboration aiming at direct searches of low-mass WIMPs and studies of $0\nu\beta\beta$ decay of $^{76}$Ge is operating two 1 kg scale p-type point-contact germanium ($p$PCGe) detectors (CDEX-1A and CDEX-1B) and one 10 kg scale $p$PCGe detector array (CDEX-10) at the China Jinping Underground Laboratory (CJPL) \cite{ref7,ref8}. Long-lived cosmogenic radionuclides, such as $^{68}$Ge, $^{68}$Ga, $^{65}$Zn have contributed clearly to the background spectra of CDEX1-A and CDEX-1B detectors  \cite{ref9,ref10,ref11,ref12}.

CDEX will proceed to a tonne-scale HPGe dark matter experiment in the future. Cosmogenic background is an important background for germanium experiments and needs to be studied in advance. This article focuses on studies of cosmogenic activation in germanium crystals to design an entire process for preparing germanium materials and detectors and ensure that the level of cosmogenic background satisfies the requirement of CDEX experiment. Section \ref{sec:2} presents the calculation process of the production rates for cosmogenic radionuclides in germanium. Several long-lived isotopes are emphatically considered in the next evaluations. The validity of this calculation was confirmed by comparing the results to data from CDEX-1B experiment in Section \ref{sec:3}. The procedure of cosmogenic background control as well as the quantitative estimations of the background level for both natural and $^{70}$Ge depleted germanium detectors are established for CDEX dark matter experiment in Section \ref{sec:4}. Based on the low background level employed in the $p$PCGe detector, the potential for new physics including solar neutrino detection and $0\nu\beta\beta$ decay detection are discussed. Finally, summaries are given in Section \ref{sec:5}.

\newpage

\section{Cosmogenic activation in germanium}\label{sec:2}
\subsection{Evaluation tools}
The cosmogenic production of radionuclides is a compound process which involves a wide energy range of particles in cosmic-ray showers and different kinds of interactions with the composition of materials. Early works of D. M. Mei et al. and E. Armengaud et al. evaluated the cosmogenic activation using TALYS, ACTIVIA, and other tools \cite{ref13, ref14, ref15}. In the present work, the evaluation was carried out with CRY \cite{ref16} and Geant4 \cite{ref17}. CRY library, version 1.7, was used to generate cosmic-ray particle showers, including six kinds of particles (neutrons, protons, muons, gammas, electrons, and pions) at three different altitudes (0, 2100, and 11300 m). We concentrate on the first four particles as the input of Geant4. Geant4 version 4.10.02 with Shielding modular physics list was used to simulate particle interactions. Shielding modular physics list contains the best-guess selection of electromagnetic and hadronic physics processes for high-energy or underground detector simulation.

\begin{figure}[!htbp]
\centering\includegraphics[width=\columnwidth]{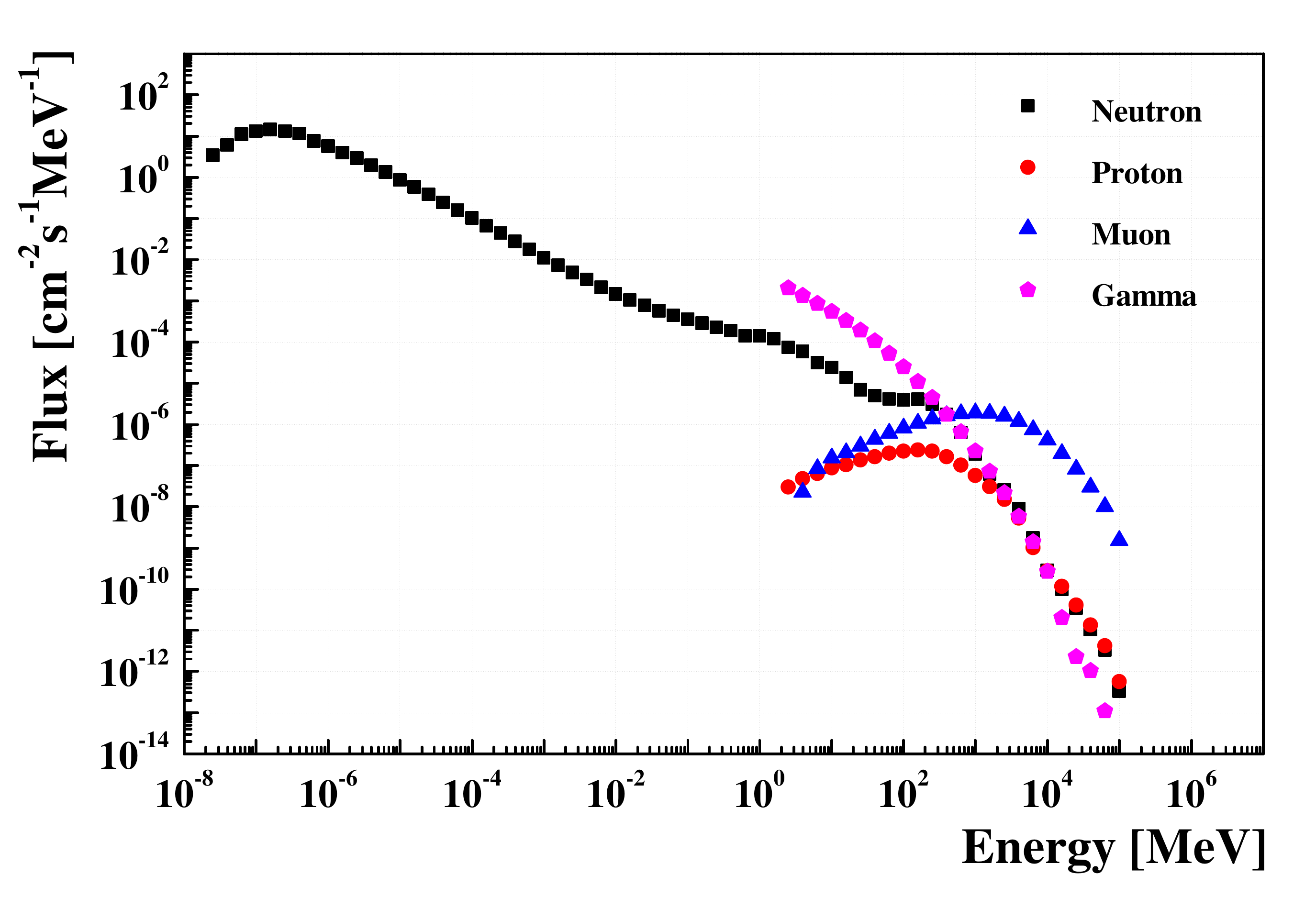}
\caption{Spectra of cosmic-ray neutrons, protons, muons, and gammas at Beijing sea level.}
\label{fig:cosmic-ray}
\end{figure}

The production rate of a radioactive isotope $i$, R$_i$, can be described as follows:
\begin{equation}
\label{eq:production_rate}
R_i = \sum_j N_j \int \Phi_k(E)\sigma_{ijk}(E)dE,
\end{equation}
where $N_j$ is the number of stable target isotope $j$, $\Phi_k$ is the flux of cosmic-ray particle $k$, and $\sigma_{ijk}$ is the production cross section of radioactive isotope $i$ produced by cosmic-ray particle $k$ on target isotope $j$. Figure \ref{fig:cosmic-ray} depicts the spectra of neutrons, protons, muons, and gammas at Beijing sea level, the integral fluxes are $2.200\times 10^{-3}$, $1.657\times 10^{-4}$, $1.182\times 10^{-2}$ and $1.682\times 10^{-2}$ cm$^{-2}$s$^{-1}$, respectively. Neutron spectrum covers from thermal neutrons to about 100 GeV neutrons, while protons, muons, and gammas distribute from 1 MeV to 100 GeV.

\subsection{Cosmogenic production rates of radionuclides}
A 1 kg germanium crystal cylinder with a height of 62 mm and a diameter of 62 mm was constructed at the center of an air box in Geant4 code. For each kind of cosmic-ray particles, primary incident particles with momentum direction were selected randomly from the corresponding spectrum generated by CRY Library and used as the input of Geant4 simulation. Secondary nuclides produced by every event were recorded, and the number of each nuclide was counted after an entire run. The cosmogenic production rate of each nuclide was calculated by adding a weight of flux value to the number of this nuclide, and the result was normalized to the unit of kg$^{-1}$day$^{-1}$. A total of 320 nuclides were counted which cover from $^{78}$Se down to $^2$H.

\begin{table*}[!htbp]
\caption{Cosmogenic production rate in natural germanium at Beijing sea level.}
\label{tab:Beijing}
\begin{tabular*}{\textwidth}{ccccp{1.8cm}<{\centering}p{1.8cm}<{\centering}p{1.8cm}<{\centering}p{1.8cm}<{\centering}p{1.6cm}<{\centering}}
\toprule
\multirow{2}{*}{Radionuclide} & \multirow{2}{*}{Half life~\cite{ref18}} & \multirow{2}{*}{Decay Mode} & \multirow{2}{*}{Daughter Nuclide} & \multicolumn{5}{c}{Production Rate (kg$^{-1}$d$^{-1}$)}\\
\cline{5-9} 
 & & & & Neutron & Proton & Muon & Gamma & Total\\
\hline
$^{68}$Ge & 270.9 d  & EC                & $^{68}$Ga & 73.30 & 5.41 & 0.31 & 4.03 & 83.05\\
$^{65}$Zn & 243.9 d  & EC or $\beta^{+}$ & $^{65}$Cu & 35.14 & 3.64 & 1.23 & 0.46 & 40.47\\
$^{63}$Ni & 101.2 yr & $\beta^{-}$       & $^{63}$Cu & 4.05  & 0.54 & 0.12 & 0.08 & 4.79\\
$^{57}$Co & 271.7 d  & EC                & $^{57}$Fe & 3.55  & 1.07 & 0.03 & 0.03 & 4.68\\
$^{60}$Co & 5.3 yr   & $\beta^{-}$ or IT & $^{60}$Ni & 1.21  & 0.22 & 0.01 & 0.01 & 1.45\\
$^{55}$Fe & 2.7 yr   & EC                & $^{55}$Mn & 3.01  & 1.05 & 0.04 & 0.05 & 4.15\\
$^{54}$Mn & 312.2 d  & EC                & $^{54}$Cr & 0.67  & 0.24 & 0.01 & 0.02 & 0.94\\
$^{49}$V  & 330.0 d  & EC                & $^{49}$Ti & 0.90  & 0.49 & 0.02 & 0.02 & 1.42\\
$^3$H     & 12.3 yr  & $\beta^{-}$       & $^3$He   & 18.33 & 4.82 & 0.33 & 0.20 & 23.68 \\
\hline
\end{tabular*}
\end{table*}

\begin{table*}[!htbp]
\caption{Cosmic-ray fluxes (cm$^{-2}$s$^{-1}$) in different areas at both sea level and 11300 m height.}
\label{tab:compaer_flux}
\begin{tabular*}{\textwidth}{p{1.8cm}<{\centering}p{2.5cm}<{\centering}p{2.5cm}<{\centering}p{2.5cm}<{\centering}p{2.5cm}<{\centering}p{2.5cm}<{\centering}p{2.5cm}<{\centering}}
\toprule
 & \multicolumn{2}{c}{\qquad Krasnoyarsk (N 56$^\circ$)} & \multicolumn{2}{c}{\qquad Strasbourg (N 49$^\circ$)} & \multicolumn{2}{c}{\qquad Beijing (N 40$^\circ$)}\\
& \qquad0 m\qquad & \qquad11300 m\qquad & \qquad0 m\qquad & \qquad11300 m\qquad & \multicolumn{1}{c}{\qquad0 m\qquad} & \multicolumn{1}{c}{\qquad11300 m\qquad} \\
\hline
Neutron & \qquad$3.400\times 10^{-3}$ & \qquad1.969 & \qquad$2.982\times 10^{-3}$ & \qquad1.415 & \qquad$2.200\times 10^{-3}$ & \qquad$8.793\times 10^{-1}$ \\
Proton & \qquad$2.169\times 10^{-4}$ & \qquad$1.453\times 10^{-1}$ &\qquad$2.043\times 10^{-4}$ & \qquad$1.044\times 10^{-1}$ & \qquad$1.657\times 10^{-4}$ & \qquad$6.457\times 10^{-2}$ \\
Muon & \qquad$1.191\times 10^{-2}$ & \qquad$8.960\times 10^{-2}$ & \qquad$1.191\times 10^{-2}$ & \qquad$8.420\times 10^{-2}$ & \qquad$1.182\times 10^{-2}$ & \qquad$7.122\times 10^{-2}$ \\
Gamma & \qquad$1.732\times 10^{-2}$ & \qquad2.755 & \qquad$1.722\times 10^{-2}$ & \qquad2.534 & \qquad$1.682\times 10^{-2}$ & \qquad2.131\\
\hline
\end{tabular*}
\end{table*}

\begin{table*}[!htbp]
\caption{Comparison of cosmogenic production rate (kg$^{-1}$d$^{-1}$) under the condition of different areas, different altitudes, and different isotopic constitution in germanium including natural and $^{70}$Ge depleted (86.6\% of $^{76}$Ge, 13.1\% of $^{74}$Ge, 0.2\% of $^{73}$Ge, and 0.1\% of $^{72}$Ge) germanium.}
\label{tab:production_rate}
\begin{tabular*}{\textwidth}{p{2cm}<{\centering}p{3cm}<{\centering}p{3cm}<{\centering}p{3cm}<{\centering}p{3cm}<{\centering}p{3cm}<{\centering}}
\toprule
\multirow{3}{*}{Radionuclide} & $^{70}$Ge depleted Ge & Natural Ge& Natural Ge& Natural Ge& Natural Ge\\
& Zelenogorsk & Zelenogorsk & Strasbourg & Beijing & Beijing  \\
& sea level & sea level & sea level & sea level & 11300 m  \\
\hline
$^{68}$Ge & 19.99 & 124.50 & 109.22 & 83.05 & 12503.79\\
$^{65}$Zn & 18.01 & 60.23  & 53.28  & 40.47 & 6366.21\\
$^{63}$Ni & 6.23  & 6.98   & 6.15   & 4.79  & 951.56\\
$^{57}$Co & 2.81  & 6.55   & 5.96   & 4.68  & 1501.11\\
$^{60}$Co & 1.86  & 1.90   & 1.85   & 1.45  & 400.37\\
$^{55}$Fe & 2.94  & 5.49   & 5.01   & 4.15  & 1521.25\\
$^{54}$Mn & 0.96  & 1.18   & 1.10   & 0.94  & 411.57\\
$^{49}$V  & 1.13  & 1.74   & 1.72   & 1.42  & 774.37\\
$^3$H     & 22.76 & 32.72  & 30.27  & 23.68 & 8691.28\\
\hline
\end{tabular*}
\end{table*}

Under cosmic-ray exposure, cosmogenic radionuclide in germanium decays when being produced. The number of one radionuclide $N$ at time $t$ can be expressed by the following formula:
\begin{equation}
\label{eq:yield}
\frac{dN}{dt} = P-\lambda N,
\end{equation}
where $P$ refers to the production rate of this radionuclide, and $\lambda$ refers to the decay constant. If $N_0$ refers to the number of the radionuclide when $t=0$, then we can rewrite Eq. (\ref{eq:yield}) as a function of $t$ as follows:
\begin{equation}
\label{eq:yield_t_N0}
N(t)=\frac{P}{\lambda}(1-e^{-\lambda t})+N_0e^{-\lambda t}.
\end{equation}

Except for $^{68}$Ge, other cosmogenic radionuclides could be removed by zone refinement and crystallization; the number of each nuclide is expected to start from a negligible level ($N_0$ = 0). When time $t$ lasts long enough, the production curve of nuclide will reach saturation, i.e., the number of the radionuclide no longer increases; the saturated value can be derived as $P/\lambda$. The process takes about 4.3 half-lives to achieve 95\% of the saturated value and 6.6 half-lives to decay to 1\%. Table~\ref{tab:Beijing} lists the major long-lived cosmogenic radionuclides produced in natural germanium at Beijing sea level; the last five columns are obtained from the Geant4 simulation. Note that gamma contributes the same order of magnitude of production rate as that of muon, and the value is higher than that in $^{68}$Ge case. 

Since the Earth's magnetic field filters low-energy particles of primary cosmic-ray, which mainly consists of galactic protons \cite{ref18}, the spectrum and flux of each cosmic-ray particle in the atmosphere varies at different latitudes. Meanwhile, cosmic-ray fluxes depend strongly on the altitude. Table \ref{tab:compaer_flux} lists the cosmic-ray fluxes calculated by CRY in different areas at both sea level and 11300 m height; these areas include the city of Zelenogorsk where isotopic separation can be carried out, the city of Strasbourg where the detector manufacturer of CANBERRA France is located, and the city of Beijing. Based on the table, the fluxes increase with increasing latitude; the fluxes of neutron, proton, and gamma at 11300 m are more than two orders of magnitude greater than that at sea level, resulting in two orders of magnitude greater production rate. Flight transportation must be avoided, and the exposure time on the Earth's surface should also be kept as short as possible to avoid bringing extra cosmic activation in the detector. Cosmogenic production rate was calculated case by case in Table \ref{tab:production_rate}.  

\section{Validation of calculation codes by CDEX-1B data}\label{sec:3}
\subsection{Exposure history of CDEX-1B}
CDEX-1B detector is a $p$PCGe detector produced by CANBERRA France in 2012. This detector was fabricated and tested in Strasbourg before airlift to Beijing in November 2013 and then transported to CJPL immediately. After 147 days in underground laboratory, the entire experimental system was established and data collection started in March 2014. Two groups of data in the early time were used for cosmogenic activation analysis, and details are listed in Table \ref{tab:data_info}. The number of each nuclide, except for $^{68}$Ge, is expected to reach a negligible level after the zone refinement for the CDEX-1B case. If $P_1$ represents the production rate during detector fabrication with a time of $t_{f}$, $P_2$ represents that during transportation from Strasbourg to CJPL with a time of $t_{t}$, and $\lambda$ refers to the decay constant, then the number of radionuclide per unit mass $N$ after $t_{f}+t_{t}$ is determined by:
\begin{equation}
N=\frac{P_1}{\lambda}(1-e^{-\lambda t_f})e^{-\lambda t_t}+\frac{P_2}{\lambda}(1-e^{-\lambda t_t}).
\end{equation}

Given the low cosmic flux in CJPL, which is $(2.0\pm 0.4)\times 10^{-10}$cm$^{-2}$s$^{-1}$ for muon \cite{ref19}, cosmogenic activation in CJPL is considered negligible. After a cooling time of $t_{c}$ in underground laboratory, the number of radionuclide changes to:
\begin{equation}
N=\frac{P_1}{\lambda}(1-e^{-\lambda t_f})e^{-\lambda (t_t+t_c)}+\frac{P_2}{\lambda}(1-e^{-\lambda t_t})e^{-\lambda t_c}.
\end{equation}

The production rate at sea level of Strasbourg can be used as $P_1$. An average transport time $t_t$ of 10 hours was assumed with the production rate $P_2$ at Beijing 11300 m height. $t_c$ is listed in the second column in Table \ref{tab:data_info}. The only unknown parameter is the time period of detector fabrication $t_f$ when the detector was under cosmic-ray exposure on the Earth's surface. Therefore, data set 1 was used to derive $t_f$ and data set 2 was used to verify this deduction.

\begin{table}[!htbp]
\caption{Time information of CDEX-1B data. Cooling time was counted from the day the detector was transported into the underground laboratory.}
\label{tab:data_info}
\centering
\begin{tabular*}{0.48\textwidth}{ccccc}
\toprule
Data & Cooling & \multirow{2}{*}{Date} & Run Time &  Live Time\\
Set & Time (day) & & (day) &  (day)\\
\hline
1 & 147 & 20140327-20140520 & 55 & 51.3 \\
2 & 331 & 20140927-20141108 & 42 & 40.4 \\
\hline
\end{tabular*}
\end{table}

\subsection{Data analysis and spectra comparison}
\begin{figure}[!htbp]
\centering\includegraphics[width=\columnwidth]{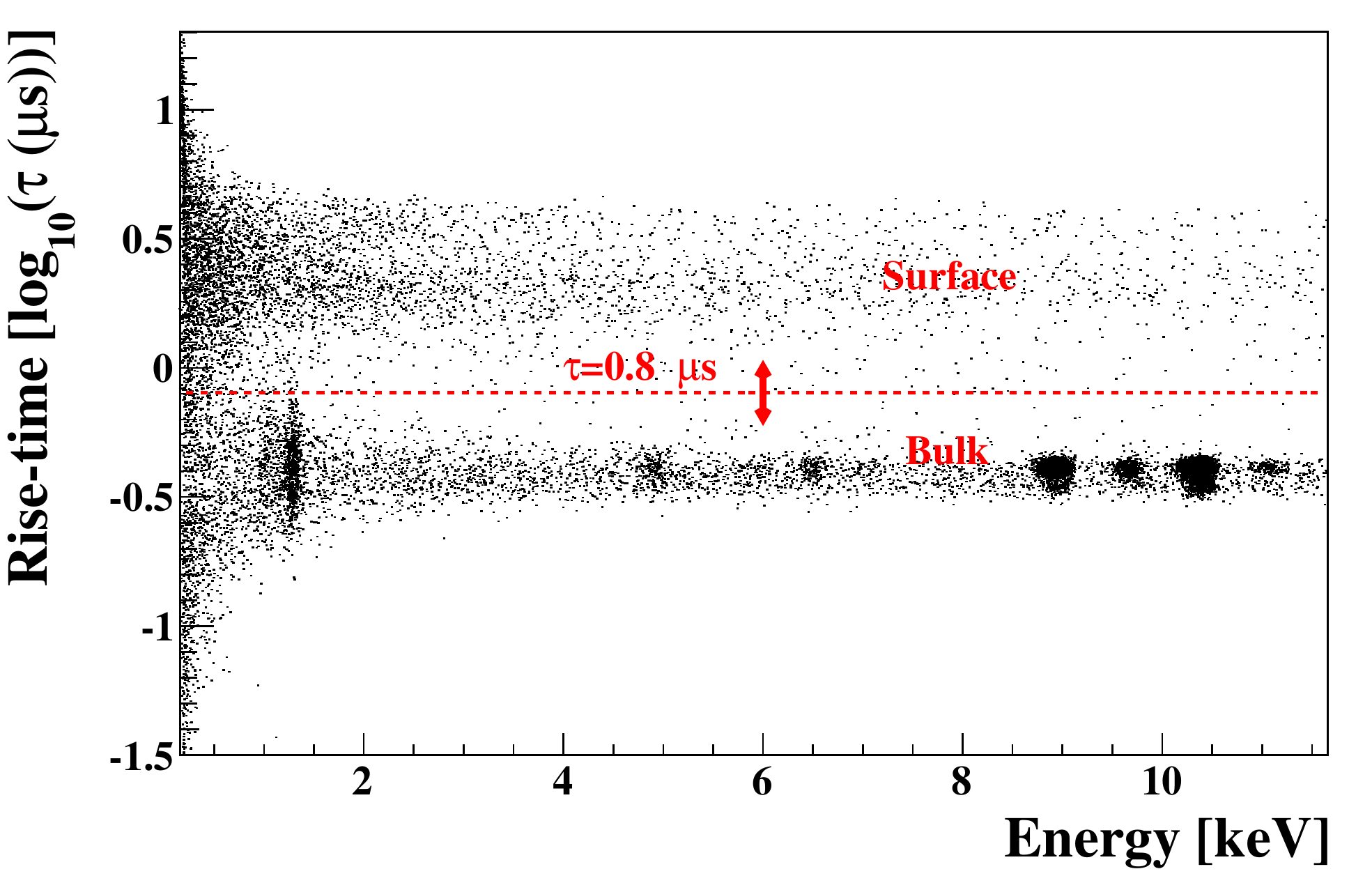}
\caption{Rise time distribution of events from the background spectrum of the CDEX-1B detector. The lower band refers to the bulk event with fast rise time, and the upper band refers to the surface events with relatively slow and dispersive rise time.}
\label{fig:RiseTime}
\end{figure}

As a p-type germanium detector, CDEX-1B contains an inactive $n^+$ layer on the germanium crystal surface. Signals originated in this layer result in partial charge collection and characteristic slow pulses, which are regarded as surface events; meanwhile, signals originated in the internal active volume with full charge collection and fast pulses are regarded as bulk events \cite{ref20,ref21,ref22}. Only bulk events can contribute to the characteristic X-ray peak of cosmogenic radionuclides; as such, the bulk spectrum was used in cosmogenic activation analysis. Figure \ref{fig:RiseTime} shows the distribution of rise time $\tau$ which is defined as the time between 5\% and 95\% of the pulse height of the signal from $p$PCGe detector. A $\tau$-cut at 0.8 $\mu s$ was set to discriminate bulk and surface events.

Two main K-shell X-ray peaks of 8.98 and 10.37 keV from $^{65}$Zn and $^{68}$Ge respectively, were compared through simulation to derive the time of $t_f$. The peaks were fitted by Gaussian functions plus a first-order polynomial. With the parameters obtained from the fitting result, peak count can be calculated by:
\begin{equation}
A=\sqrt{2\pi}\cdot\sigma\cdot H / W,
\end{equation}
where $A$ refers to the peak count, $\sigma$ refers to the sigma of the fitted Gaussian function, $H$ refers to the peak height, and $W$ refers to the energy bin size. By comparing the peak count between measurement and simulation, the value of $t_f$ was derived as 270 days, which means that the process of detector fabrication without cosmogenic control corresponds to an exposure time of a few hundred days.  

\begin{figure}[!htbp]
\centering\subfigure[]{\includegraphics[width=\columnwidth]{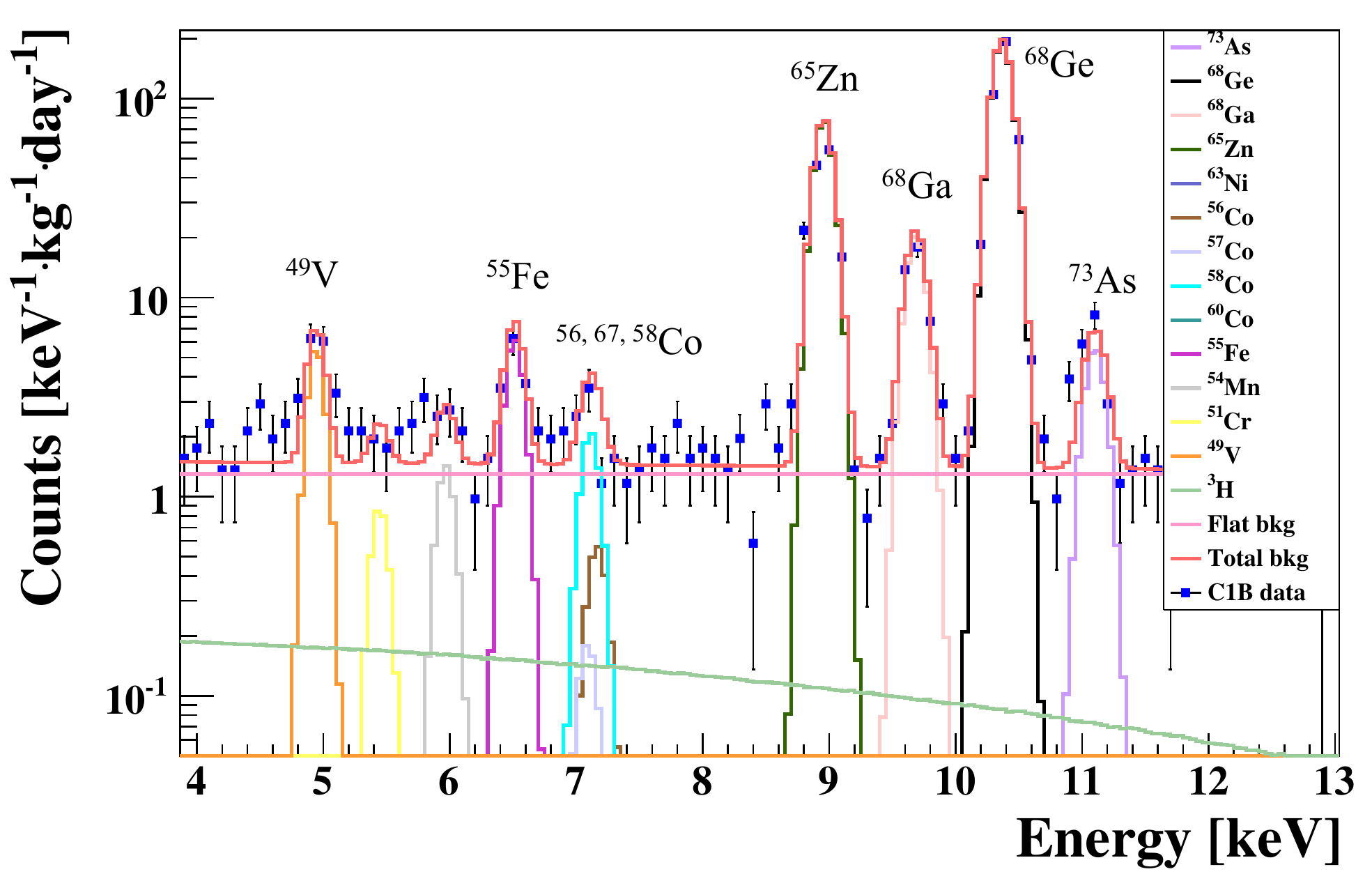}}
\centering\subfigure[]{\includegraphics[width=\columnwidth]{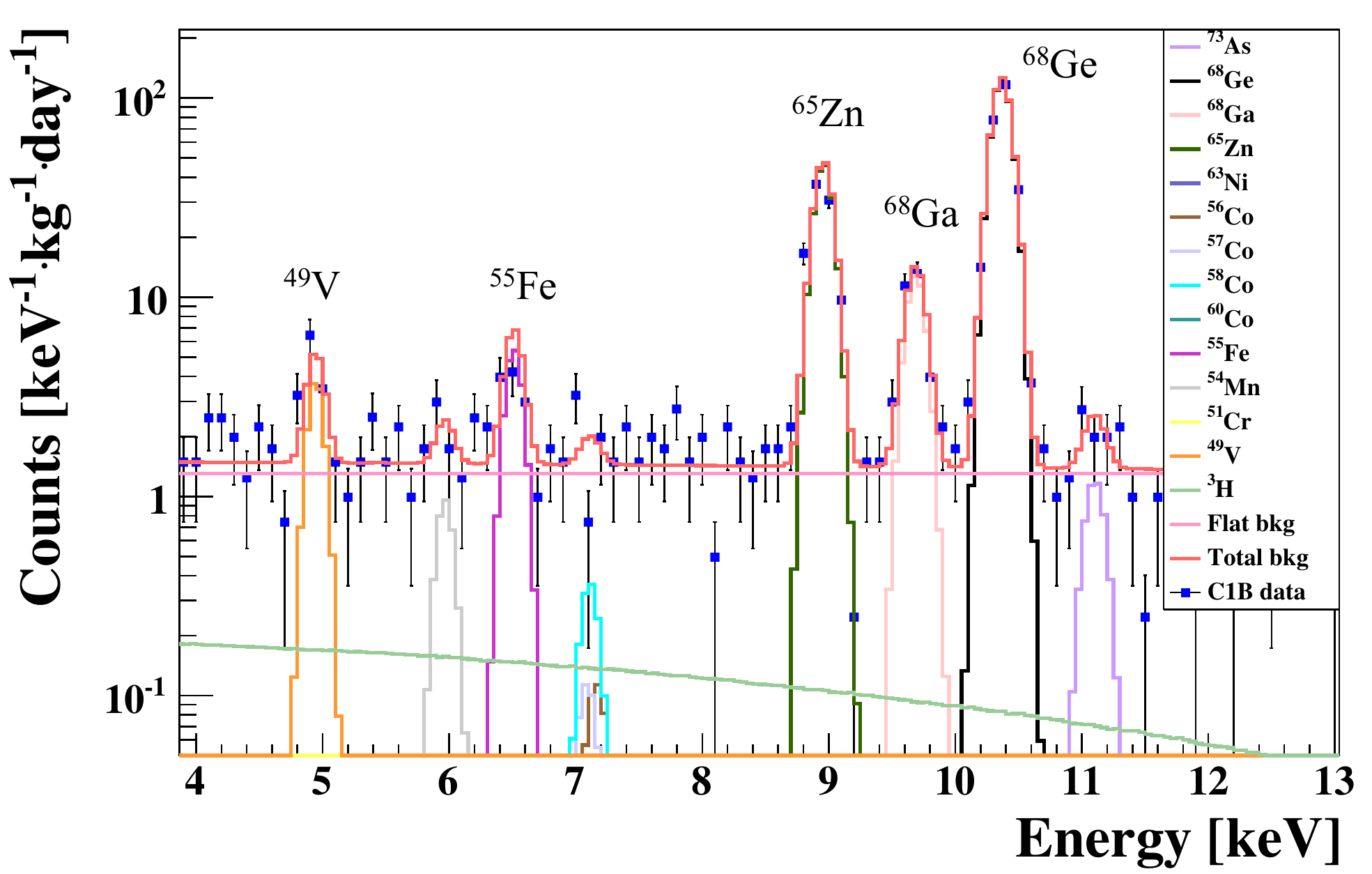}}
\caption{Spectra comparison between simulation and measurement of data set 1 (a) and data set 2 (b). Data are shown as blue dots with statistical uncertainties. Spectrum from each nuclide is shown in different colors, and the sum of all spectra is shown as red lines.}
\label{fig:fit_dataset1_2}
\end{figure}

Figure~\ref{fig:fit_dataset1_2} depicts the comparison between simulated spectra based on the calculation above and the measured data. A flat background caused by high-energy gammas was assumed and fitted within the energy range of 4 keV to 12 keV. Peaks from other radionuclides, except $^{65}$Zn and $^{68}$Ge, are used for tests. Given that the half-life of $^{68}$Ga is only 67.7 minutes, the K-shell X-ray peak of $^{68}$Ga nuclide in the spectrum originates from $^{68}$Ge electron capture decay. The resolution function used in simulation spectra is:
\begin{equation}
\sigma(keV)=0.0132\sqrt{E(keV)} + 0.0335.
\end{equation}

As shown in Figure~\ref{fig:fit_dataset1_2} and Table \ref{tab:calculation_result}, several main K-shell X-ray peaks on the spectrum from $^{49}$V, $^{55}$Fe, $^{65}$Zn, $^{68}$Ga, and $^{68}$Ge consist of data within statistical uncertainty. The calculation of $^3$H might be less accurate because of lack of characteristic X-ray for comparison. Considering the production rate of $^3$H calculated in reference \cite{ref13,ref14,ref23,ref24}, these results differ from one another by a factor of two. Figure \ref{fig:fit_dataset1_2} shows that the background level of $^3$H is about 10$^{-1}$ cpkkd for commercial natural germanium detector. From this comparison, the simulation procedure setup to calculate the cosmogenic background in CDEX-1B was validated. The production rates of radionuclides derived from this simulation method were further used to evaluate the cosmogenic background level for future experiments.

\begin{table}[!htbp]
\caption{Comparison between calculation and the fit result of experimental peak count of data set, 1 with $t_f=270$ days, $t_t=10$ hours, $t_c=147$ days, and run time of 55 days.}
\label{tab:calculation_result}
\centering
\begin{tabular*}{0.48\textwidth}{cccc}
\toprule
\multirow{2}{*}{Radionuclide} & \quad Measured & \quad Measured & \quad Simulated \\
                   & \quad Energy (keV) & \quad Peak Count & \quad Peak Count \\
\hline
$^{68}$Ge       & \quad 10.38$\pm$0.01 & \quad 1919.4$\pm$68.8 & \quad 1918.7\\
$^{68}$Ga       & \quad 9.68$\pm$0.01  & \quad 190.8$\pm$23.9 & \quad 203.3\\
$^{65}$Zn       & \quad 8.95$\pm$0.01  & \quad 685.8$\pm$38.7 & \quad 689.4\\
$^{56,57,58}$Co & \quad 7.14$\pm$0.02  & \quad 21.4$\pm$17.2 & \quad 22.2\\
$^{55}$Fe       & \quad 6.50$\pm$0.02  & \quad 44.1$\pm$16.9 & \quad 55.4\\
$^{49}$V        & \quad 4.95$\pm$0.02  & \quad 52.3$\pm$19.6 & \quad 46.5\\
\hline
\end{tabular*}
\end{table}

\section{Cosmogenic background in the future tonne-scale CDEX experiment}\label{sec:4}

\begin{table*}[!htbp]
\caption{Controlled process of natural germanium detector before moving to CJPL.}
\label{tab:naturalGe}
\begin{tabular*}{\textwidth}{p{5.8cm}p{1.2cm}<{\centering}p{3.5cm}p{7cm}}
\toprule
\multirow{2}{*}{Event} & Event & \multirow{2}{*}{Shielding condition} & \multirow{2}{*}{Comments}\\
 & duration &  & \\
\hline
Prepare natural Ge and transport to U.S.& Flexible & No shielding & Saturation for all radionuclides \\
Detector fabrication & 1 month & Temporary underground & Start to calculate the cosmogenic yield of \\
 & & storage & each radionuclide except for $^{68}$Ge \\
Ship and ground transport to CJPL & 1 month & Shielding container& Reduce consmogenic activation by a factor of 10 \\
\hline
\end{tabular*}
\end{table*}

\begin{table*}[!htbp]
\caption{Controlled process of $^{70}$Ge depleted germanium detector before moving to CJPL.}
\label{tab:depletedGe}
\begin{tabular*}{\textwidth}{p{5.8cm}p{1.2cm}<{\centering}p{3.5cm}p{7cm}}
\toprule
\multirow{2}{*}{Event} & Event duration & \multirow{2}{*}{Shielding condition} & \multirow{2}{*}{Comments} \\
\hline
 $^{70}$Ge depletion in the form of GeF$_4$ gas & 40 days & No shielding & Start to calculate the cosmogenic yield of $^{68}$Ge \\
Storage of GeO$_2$ powder & 1 month & Underground or under shield & Wait for transportation \\
Ground transport of GeO$_2$ to CJPL & 15 days & Shielding container & Reduce consmogenic activation by a factor of 10 \\
Convert GeO$_2$ to Ge metal & Flexible & Underground & Consmogenic activation in CJPL can be negligible \\
Transport Ge metal to U.S.& 1 month & Shielding container & Reduce consmogenic activation by a factor of 10 \\
Detector fabrication & 1 month & Temporary underground & Start to calculate the cosmogenic yield of each \\
 &  & storage & radionuclide except for $^{68}$Ge \\
Ship and ground transport to CJPL & 1 month & Shielding container& Reduce consmogenic activation by a factor of 10 \\
\hline
\end{tabular*}
\end{table*}

In the future tonne-scale CDEX experiment, the HPGe detector array will be immersed in a large liquid nitrogen shielding system to reduce the background level. When the external background from both the environment and materials are expected to reach a quite low level, the cosmogenic background from the internal germanium crystal will be highlighted.  

\subsection{Controlled germanium processing and detector fabrication}
\subsubsection{Natural germanium}
\label{sec:naturalGe}
The main process in the time flow of natural germanium detector before finally storing at CJPL is summarized in Table \ref{tab:naturalGe}. Without special treatment or protection, the number of each cosmogenic radionuclide in natural germanium is expected to reach saturation. During detector fabrication, radionuclides, except for $^{68}$Ge, can be removed effectively by zone refinement and crystallization. A total working time of 1 month with 8 hours per day is assumed, and the detector can be stored in a shallow underground laboratory for the rest of the time. The activation occurs while germanium crystal is above ground. If the detector is fabricated in the United States of America, then a one-way transport time of at least 1 month is needed. During transportation, properly designed shields can reduce the cosmogenic activation by a factor of 10 according to the calculations by Barabanov et al. \cite{ref25,ref26}. 

The background can be further reduced by moving the complete detector fabrication process to underground laboratory. Natural germanium material can be directly transported to CJPL. Thus, only two cosmogenic radionuclides, namely, $^{68}$Ge and its daughter nuclide $^{68}$Ga, are left to be considered. $^{68}$Ge is mainly produced by cosmic-ray on stable isotope $^{70}$Ge.

\subsubsection{$^{70}$Ge depleted germanium}
\label{sec:depletedGe}
The authors took the isotopic separation carried out in Electrochemical Plant (ECP) in Zelenogorsk, Russia as an example. A summary of all the steps in the procedure are shown in Table \ref{tab:depletedGe}. Germanium is depleted in the form of stable GeF$_4$ gas and then converted to GeO$_2$ powder for storage after achieving the required isotope abundance. The cosmogenic activation starts right after the GeF$_4$ gas leaves the centrifuge \cite{ref27}. Separation for 40 days is assumed for producing a batch of GeO$_2$ powder, and the activation is effectively reduced when stored under shield. Ground transportation of 15 days is needed from Zelenogorsk to CJPL, where GeO$_2$ is deoxidized into metal. The subsequent process for detector fabrication is the same as that described in section \ref{sec:naturalGe}, that is, 1 month for transport of $^{70}$Ge depleted germanium metal from CJPL to U.S. company, 1 month for detector fabrication in the U.S., and 1 month for detector transportation from U.S. company to CJPL. The cosmogenic activation in underground laboratory is considered to reach a negligible level, and all transportations are under shield with the reduction of activation by a factor of 10.

Compared with natural germanium, the amounts of $^{68}$Ge and $^{68}$Ga in $^{70}$Ge depleted germanium material can be effectively suppressed. If the process of crystal growth and detector fabrication can be carried out in underground laboratory, then only the first three rows in Table~\ref{tab:depletedGe} need to be considered. Quantitative analysis of cosmogenic background in natural and $^{70}$Ge depleted germanium will be discussed in the following sections.

\subsection{Cosmogenic background of HPGe detectors for dark matter detection}
\subsubsection{Natural germanium}

\begin{figure}[!htbp]
\centering\includegraphics[width=\columnwidth]{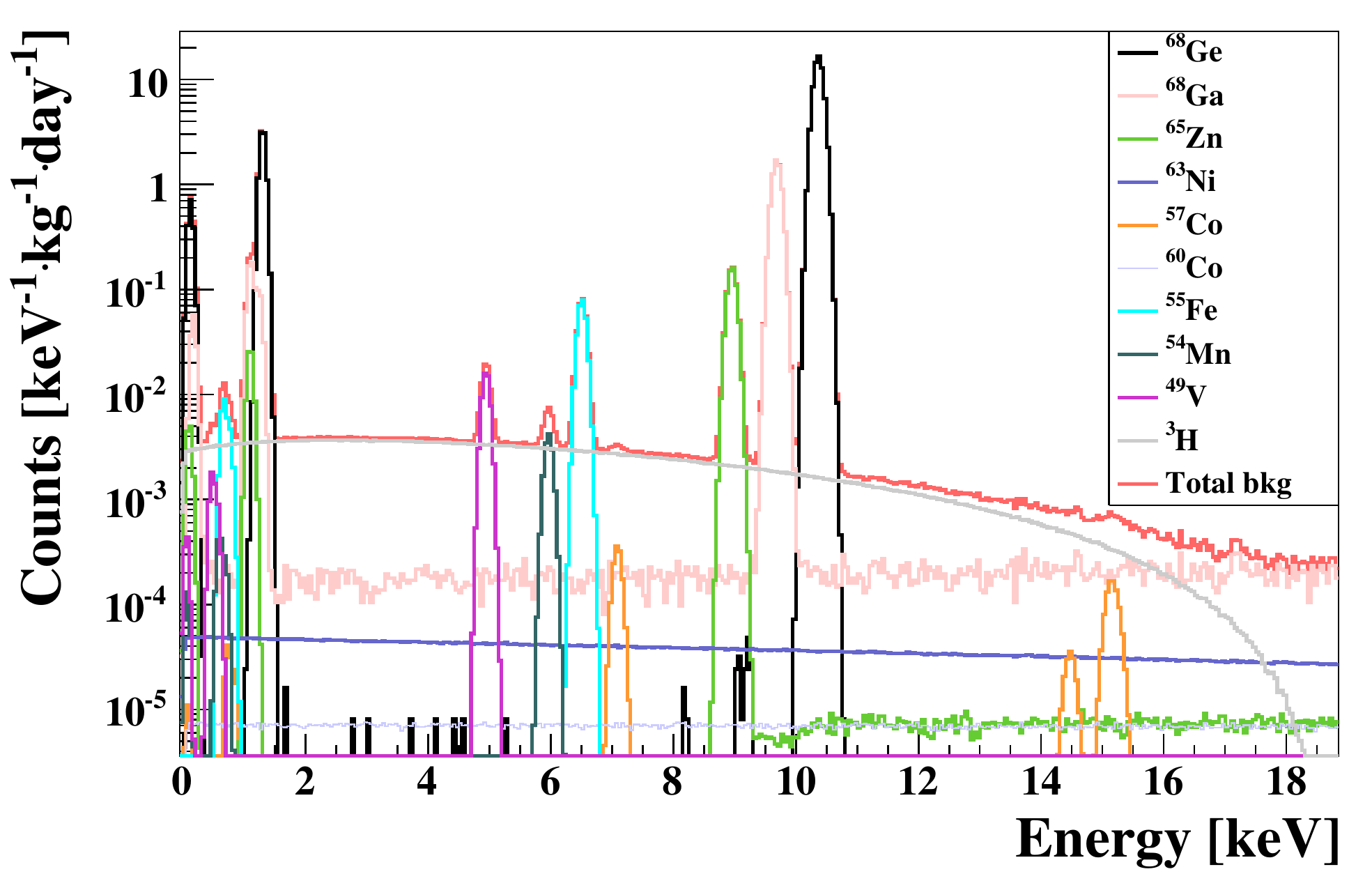}
\caption{Cosmogenic background spectra at the range of interest in natural germanium with 3 years of cooling time. Lines in different colors refer to the spectra of different radionuclides. The exposure time includes 1 month of detector fabrication and 1 month of transportation.}
\label{fig:natural_3years}
\end{figure}

Based on the process described in section \ref{sec:naturalGe}, the cosmogenic background in the dark matter detection region of interest with 3 years of cooling time at CJPL is shown in Figure \ref{fig:natural_3years}. The continuous $\beta ^-$ spectrum of $^3$H with an end point of 18.6 keV dominates the background at 2 $\sim$ 4 keV range, which is about 3$\times$10$^{-3}$ cpkkd. Since the number of $^{68}$Ge in natural germanium is expected to reach saturation, the L-shell and M-shell X-ray of $^{68}$Ge and $^{68}$Ga mainly contribute the background around and below 1 keV, which is three orders of magnitude higher than the continuous background.

\begin{figure}[!htbp]
\centering\subfigure[]{\includegraphics[width=0.49\columnwidth]{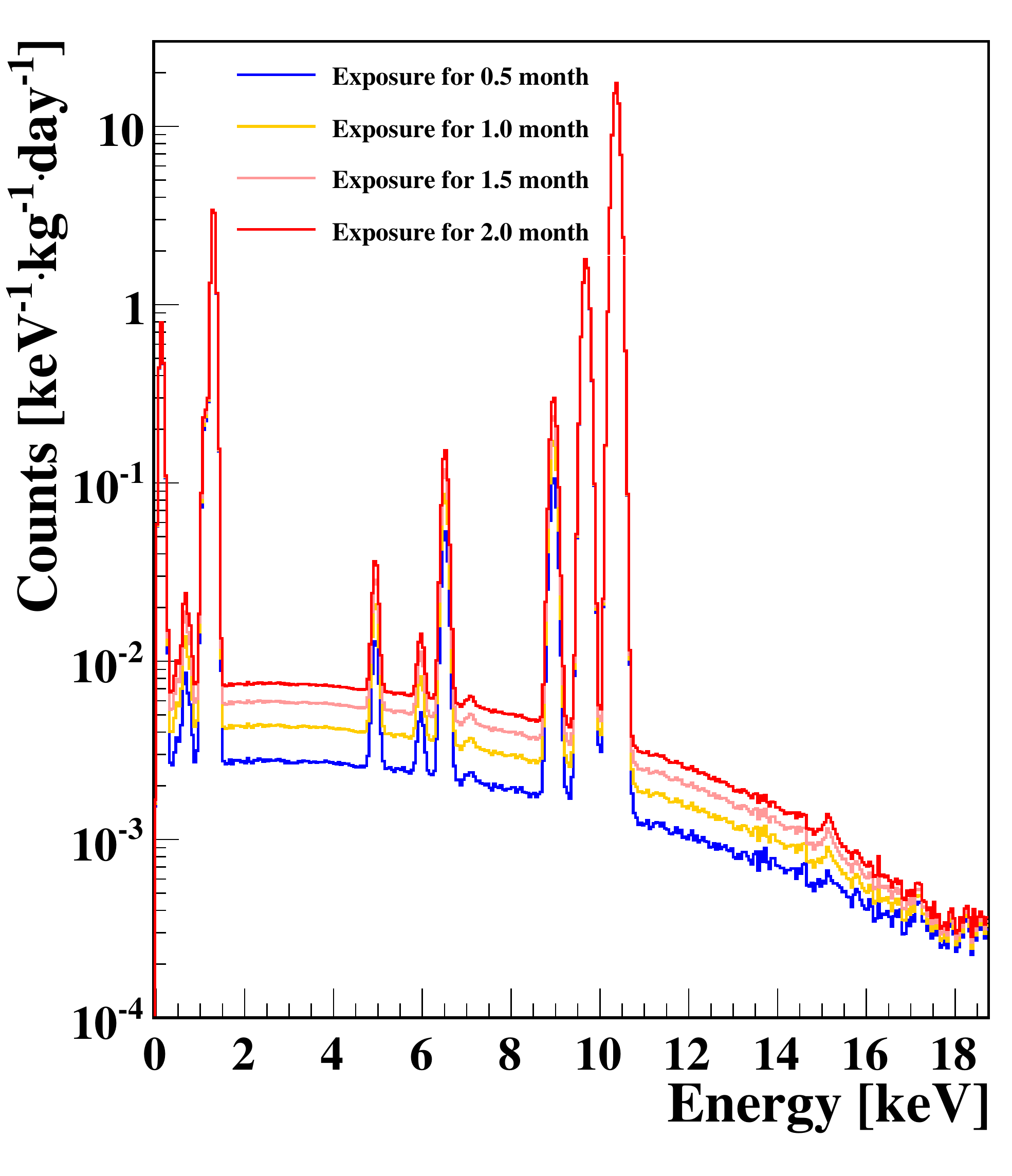}}
\centering\subfigure[]{\includegraphics[width=0.49\columnwidth]{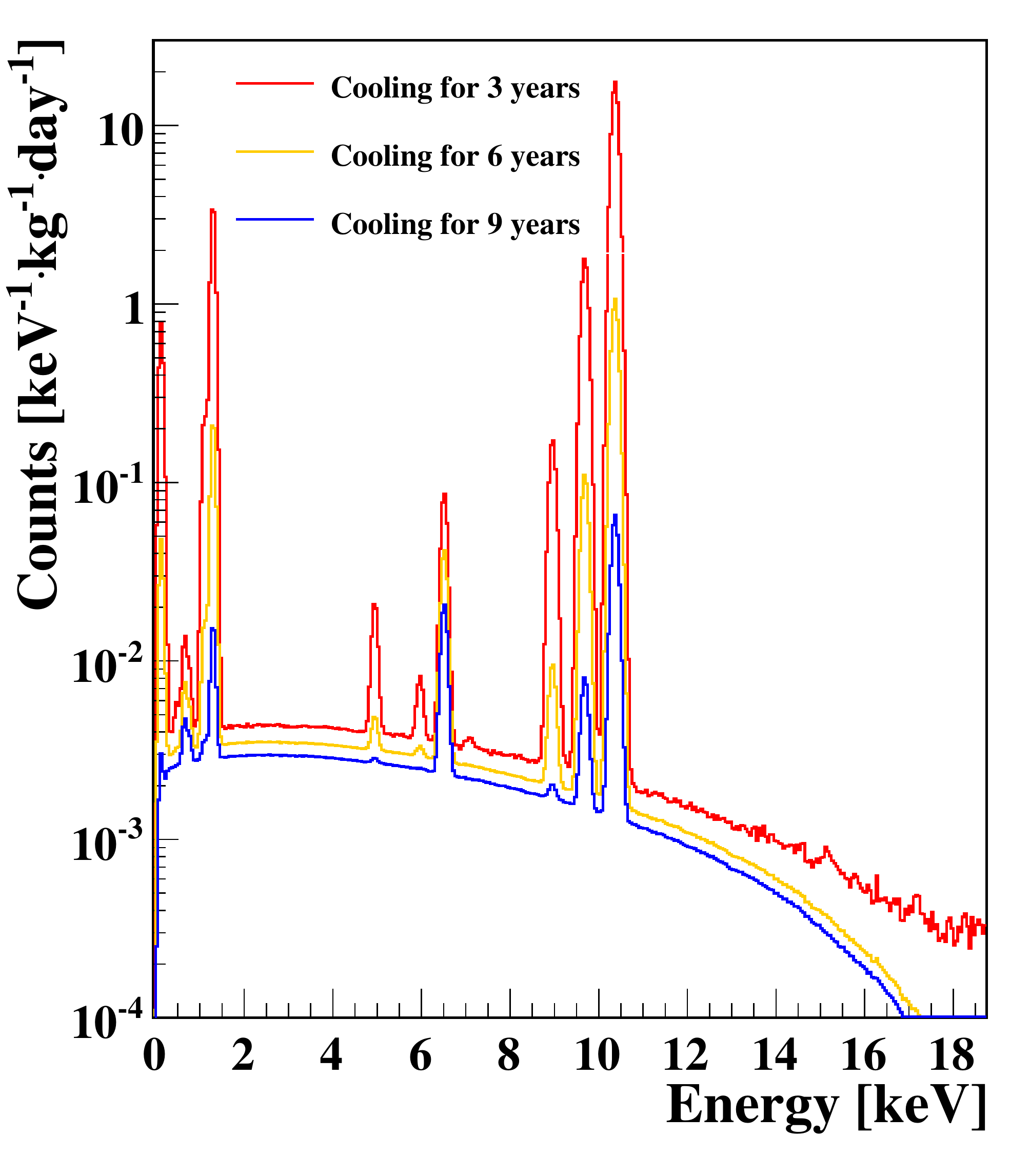}}
\caption{The background level variations with different detector fabrication times with 3 years cooling (a) and different cooling times with 1 month exposure (b).}
\label{fig:natural_col_exp}
\end{figure}

With 3 years cooling at CJPL, changing the time of detector fabrication from half month to 2 months only brings a difference of two to three times within 2 $\sim$ 4 keV range, as shown in Figure \ref{fig:natural_col_exp} (a). Due to the half-life of 12.3 years, the contribution of $^3$H does not change substantially even when the cooling time is extended to 9 years. However, with a shorter half-life, the characteristic X-ray peaks from other radionuclides around 1 keV are significantly decayed by about two orders of magnitude, as shown in Figure \ref{fig:natural_col_exp} (b).

\begin{figure}[!htbp]
\centering\subfigure[]{\includegraphics[width=0.49\columnwidth]{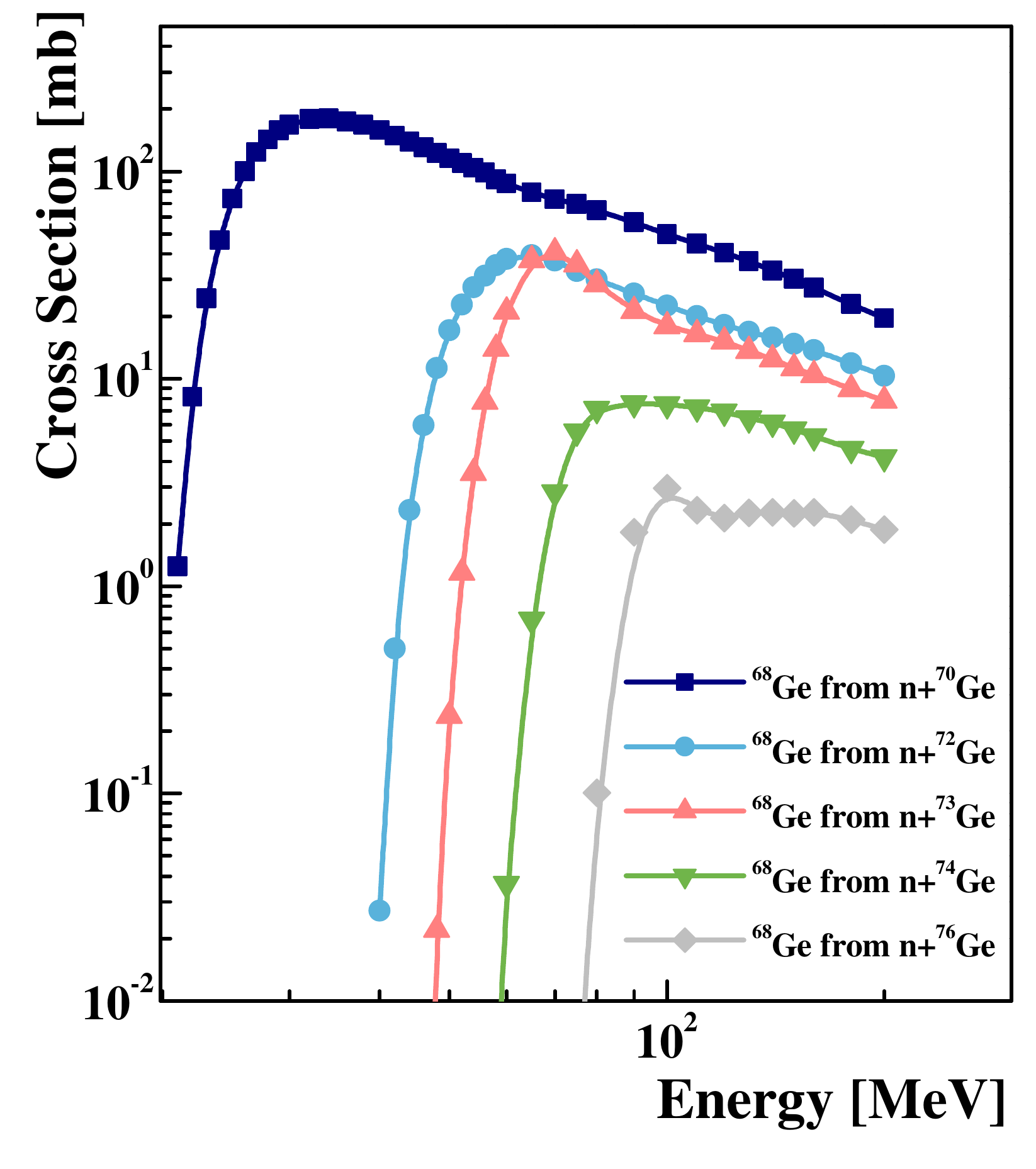}}
\centering\subfigure[]{\includegraphics[width=0.49\columnwidth]{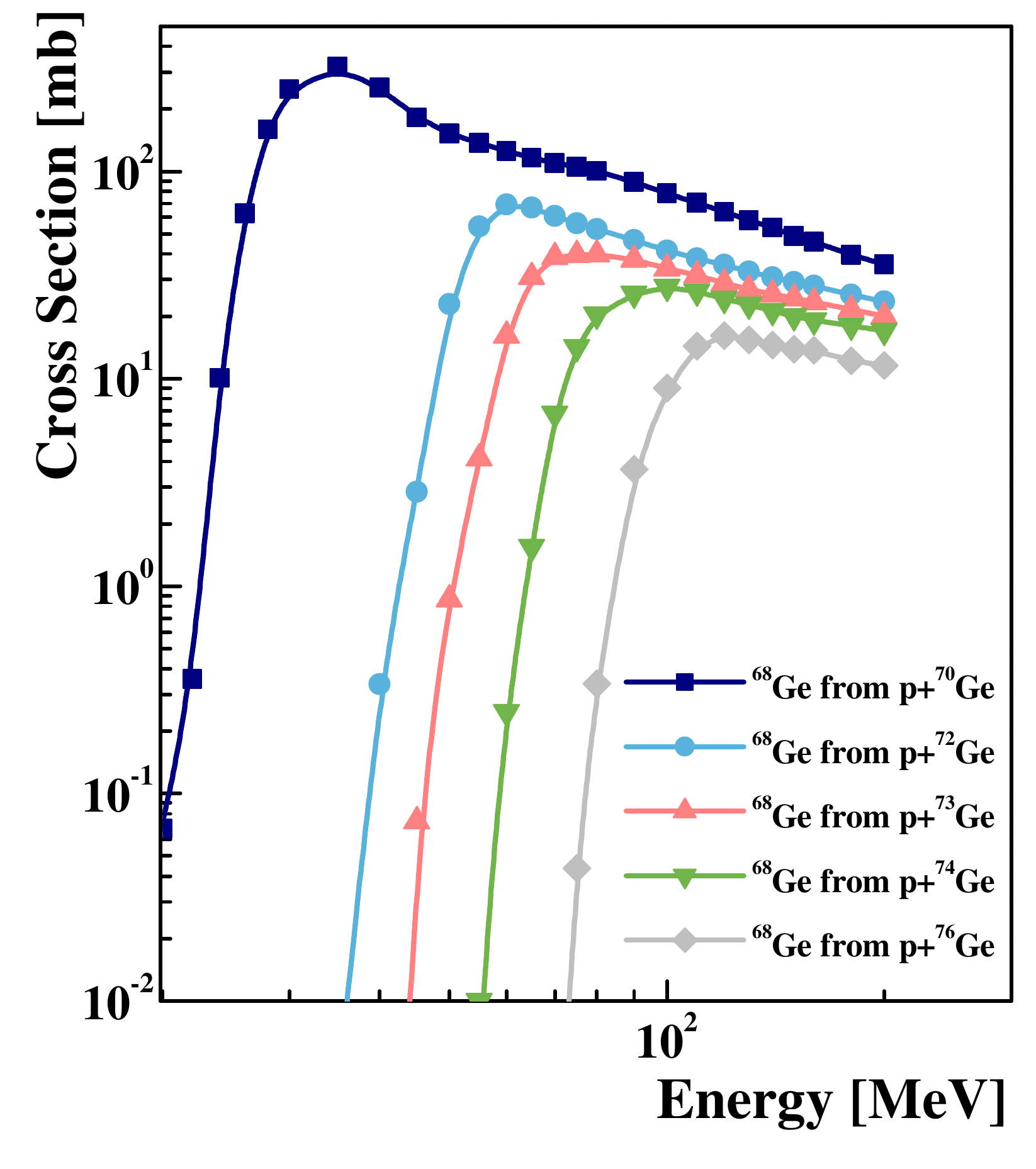}}
\caption{Total cross section of $^{68}$Ge production by neutron and proton on stable isotopes of germanium.}
\label{fig:68Ge_cross_section}
\end{figure}

When the background level reduces to below 10$^{-2}$ cpkkd and the characteristic X-ray peaks are still 1 cpkkd, the error induced by peak subtraction would be huge. In addition to extending the underground cooling time, depletion of $^{70}$Ge can help reduce this error and further provide stricter restrictions on direct dark matter detection around and below 1 keV. As the main contributor of characteristic X-ray peaks, $^{68}$Ge can be effectively reduced by  $^{70}$Ge depletion, as shown in Figure \ref{fig:68Ge_cross_section} which contains the total cross section of $^{68}$Ge production by neutron and proton on the stable isotopes of germanium from TENDL-2015 data library. 

\subsubsection{$^{70}$Ge depleted germanium}
\begin{figure}[!htbp]
\centering\includegraphics[width=\columnwidth]{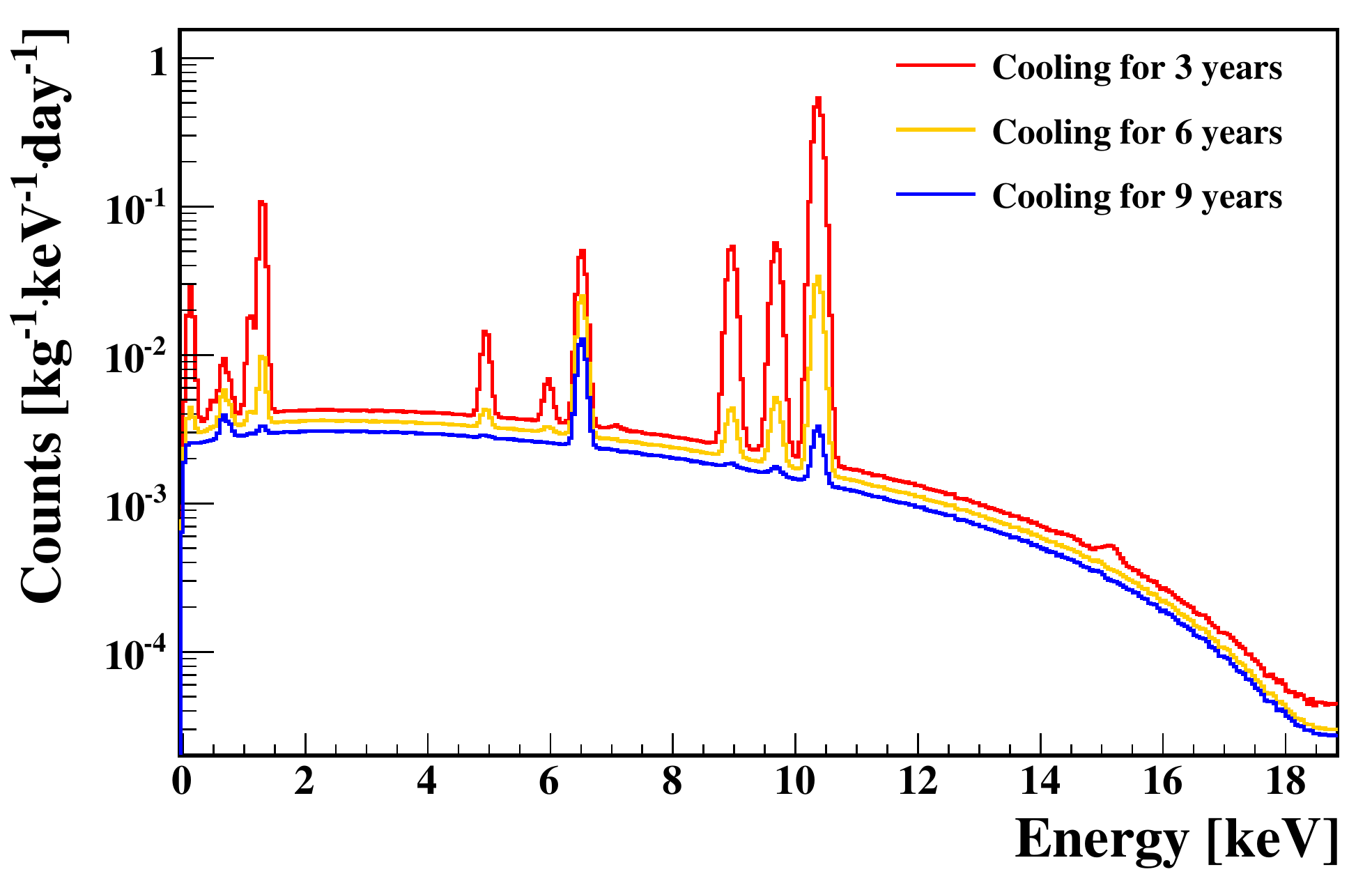}
\caption{Cosmogenic background spectra at low-energy range in $^{70}$Ge depleted germanium with different cooling times. The exposure time includes 40 days of $^{70}$Ge depletion, 1 month of detector fabrication, and 2.5 months of transportation in total.}
\label{fig:depleted_col_low}
\end{figure}

\begin{figure}[!htbp]
\centering\subfigure[]{\includegraphics[width=0.49\columnwidth]{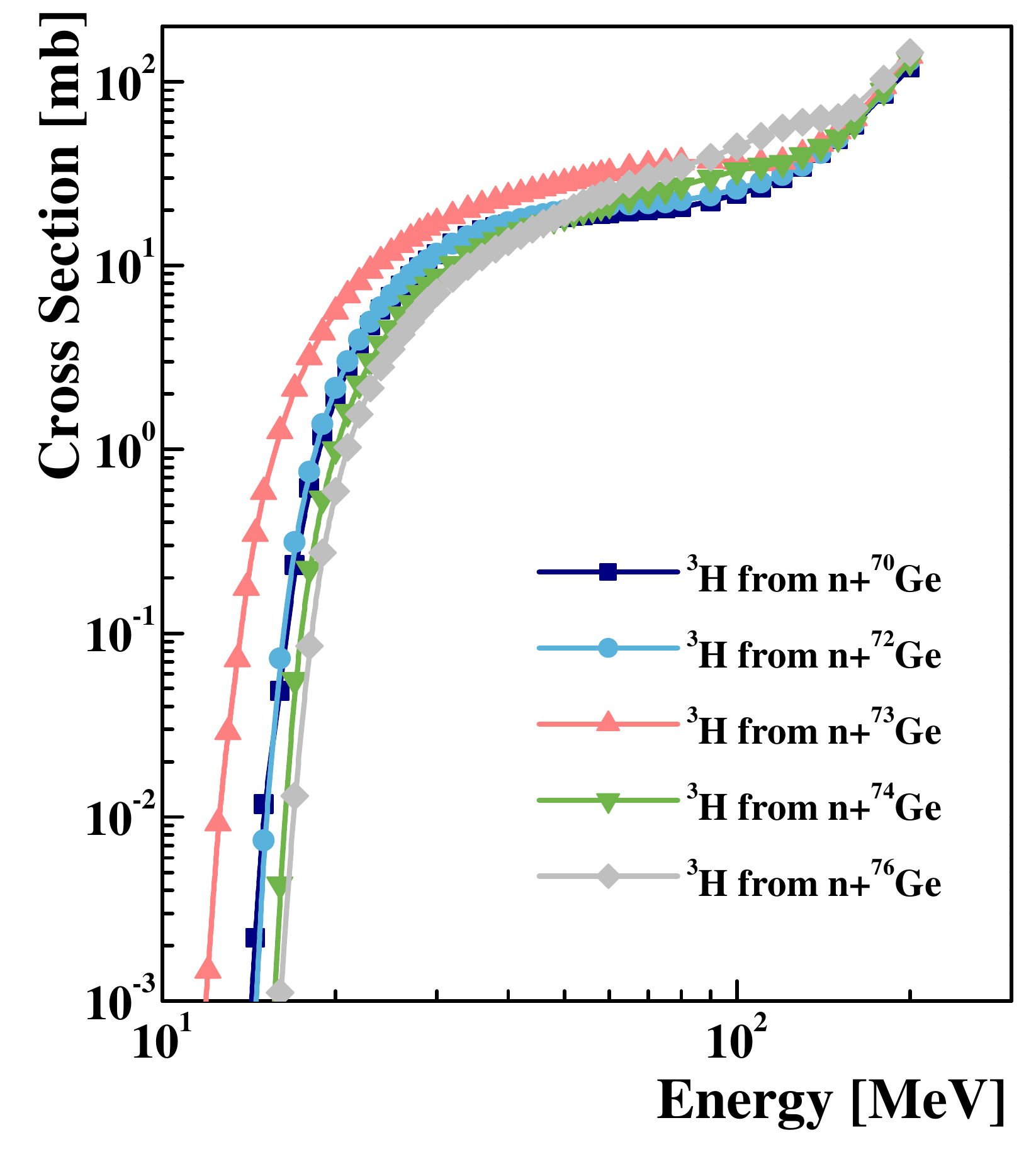}}
\centering\subfigure[]{\includegraphics[width=0.49\columnwidth]{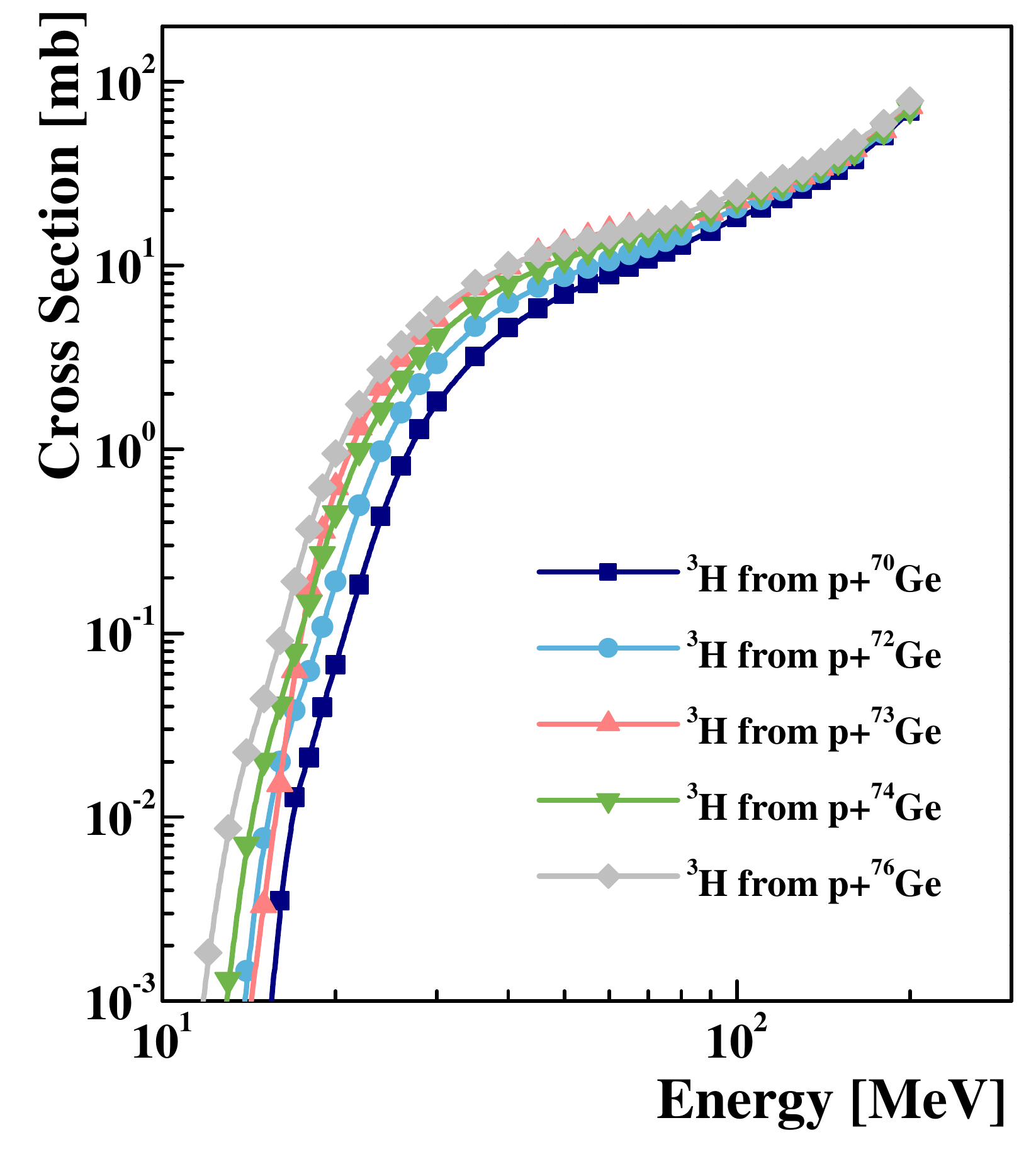}}
\caption{Total cross section of $^3$H production by neutron and proton on stable isotopes of Ge.}
\label{fig:3H_cross_section}
\end{figure}

The simulation was carried out on $^{70}$Ge depleted germanium (86.6\% of $^{76}$Ge, 13.1\% of $^{74}$Ge, 0.2\% of $^{73}$Ge, and 0.1\% of $^{72}$Ge) according to the process described in section \ref{sec:depletedGe}, the background level with different cooling times are shown in Figure \ref{fig:depleted_col_low}. Compared with Figure~\ref{fig:natural_col_exp} (b), the intensity of characteristic X-ray peaks generated by $^{68}$Ge and $^{ 68}$Ga are significantly reduced by more than one order of magnitude with the same cooling time. However, $^3$H $\beta ^-$ decay contributes the same level of background within the 2 $\sim$ 4 keV range whether deplete $^{70}$Ge or not. This same level of contribution can be understood from the total cross section of $^3$H production on stable Ge isotopes shown in Figure~\ref{fig:3H_cross_section}, with little difference between isotopes.

\subsubsection{Underground crystal growth and detector fabrication}
\begin{figure}[!htbp]
\centering\subfigure[]{\includegraphics[width=0.49\columnwidth]{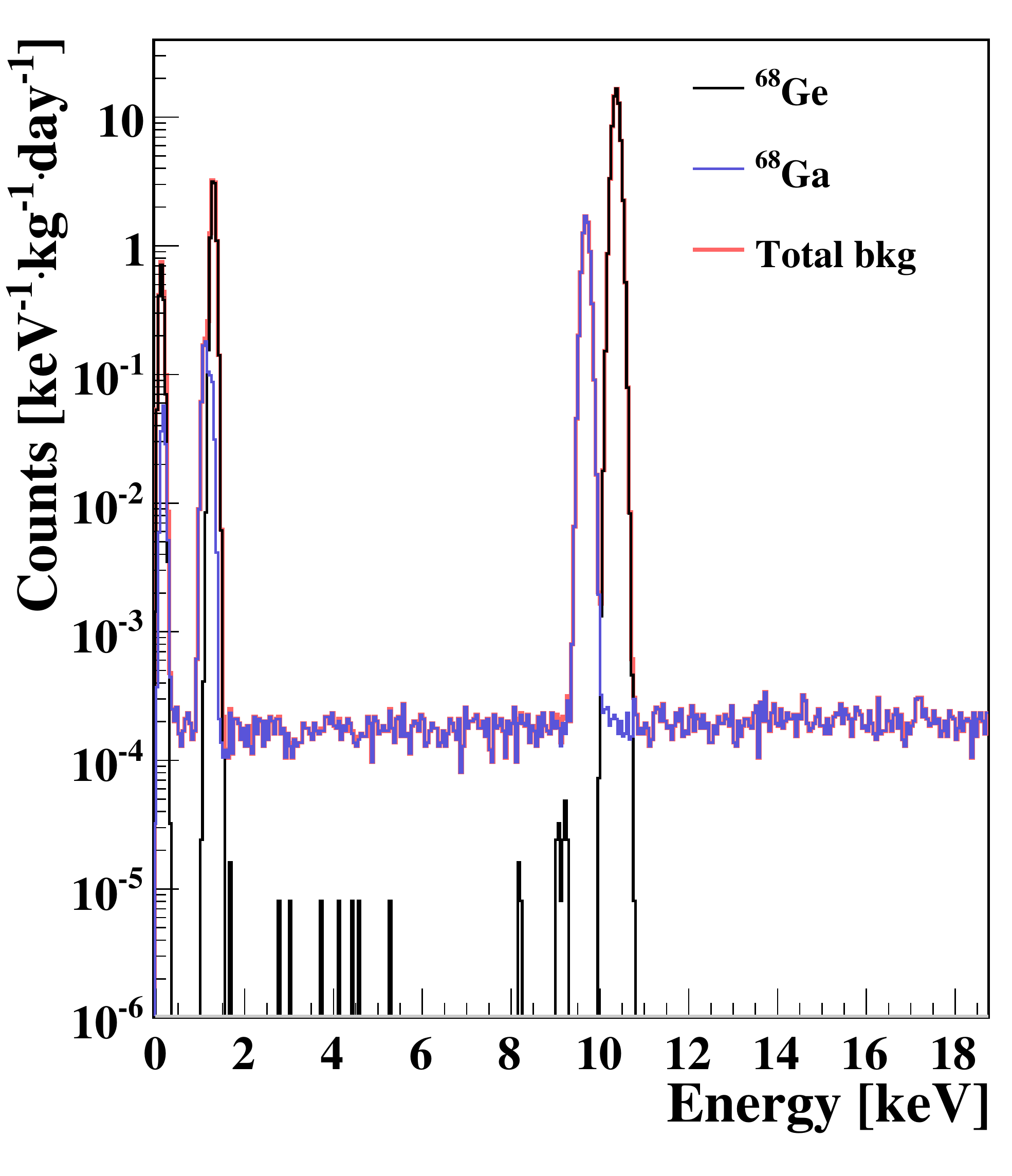}}
\centering\subfigure[]{\includegraphics[width=0.49\columnwidth]{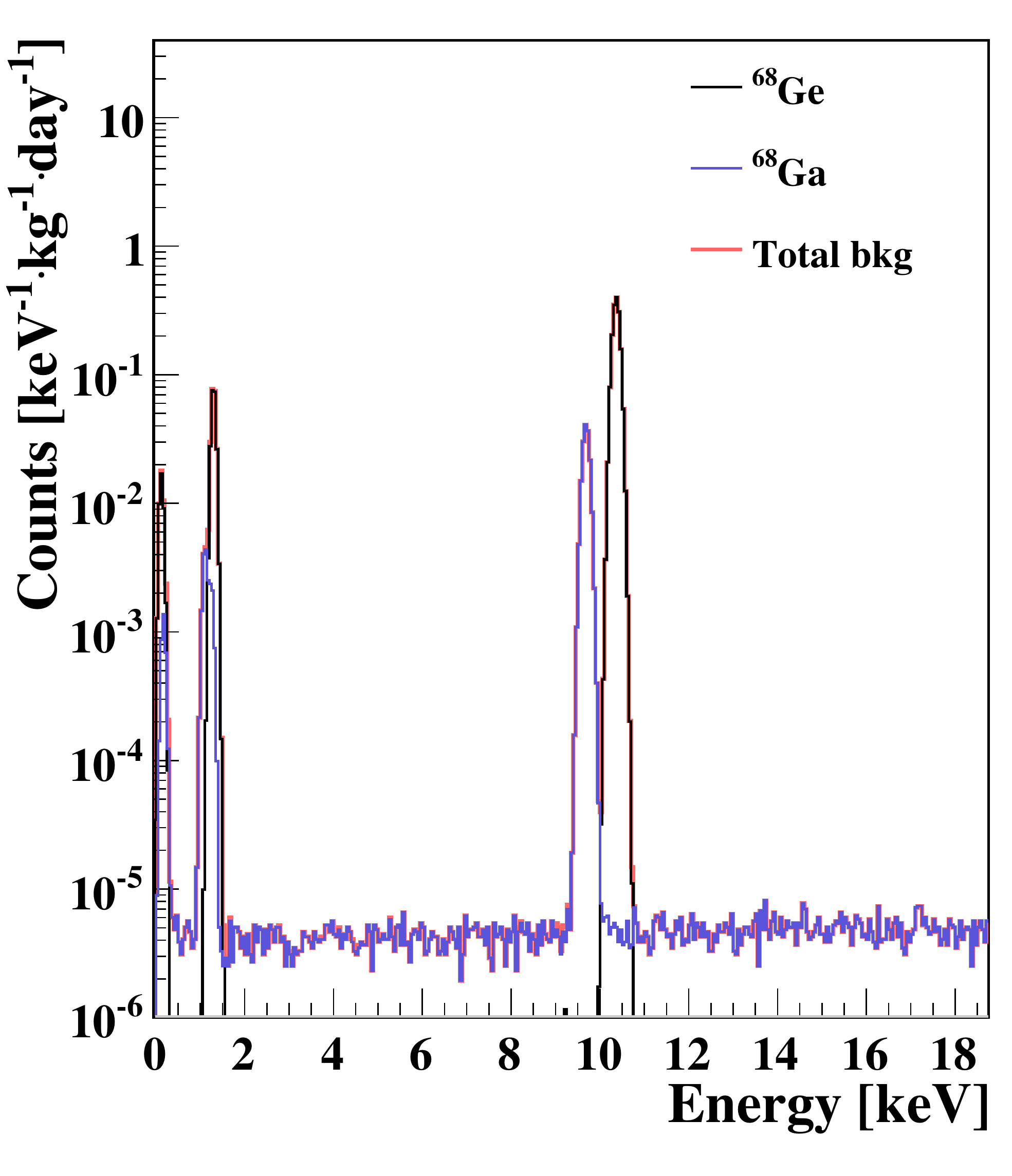}}
\caption{Background level of natural (a) and $^{70}$Ge depleted (b) germanium with underground crystal growth and detector fabrication.}
\label{fig:underground}
\end{figure}

With underground crystal growth and detector fabrication, the background could be decreased to quite a low level within a reasonable cooling time of 3 years. The cosmogenic spectra of natural and $^{70}$Ge depleted germanium are shown in Figure \ref{fig:underground} (a) and (b), respectively. For natural germanium, the amount of $^{68}$Ge is calculated with saturation number; for $^{70}$Ge depleted germanium, the amount of $^{68}$Ge is calculated based on the exposure time of 40 days of depletion plus 15 days of transportation. 

Except for $^{68}$Ge and $^{68}$Ga, the contribution of $^3$H and other radionuclides could dramatically decrease to negligible level. The continuous spectrum at 2 $\sim$ 4 keV can even reduce to less than 10$^{-5}$ cpkkd in $^{70}$Ge depleted germanium. At the same time, the characteristic X-ray peaks are reduced by more than one order of magnitude. With such a low background level employed in the $p$PCGe detector, it provides us the opportunity to further decrease the background level and improve the detection sensitivity to touch the solar neutrino floor for CDEX tonne-scale dark matter experiment. The ultra-low background level in the low-energy region (0 $\sim$ 20 keV) and high-energy region ( $\sim$ 2 MeV) also provides us the potential for opening new physical windows, including solar neutrino and $0\nu\beta\beta$ decay detection.

\subsection{Potential for new physics}
\subsubsection{Solar neutrino detection}
\begin{figure}[!htbp]
\centering\subfigure[]{\includegraphics[width=0.49\columnwidth]{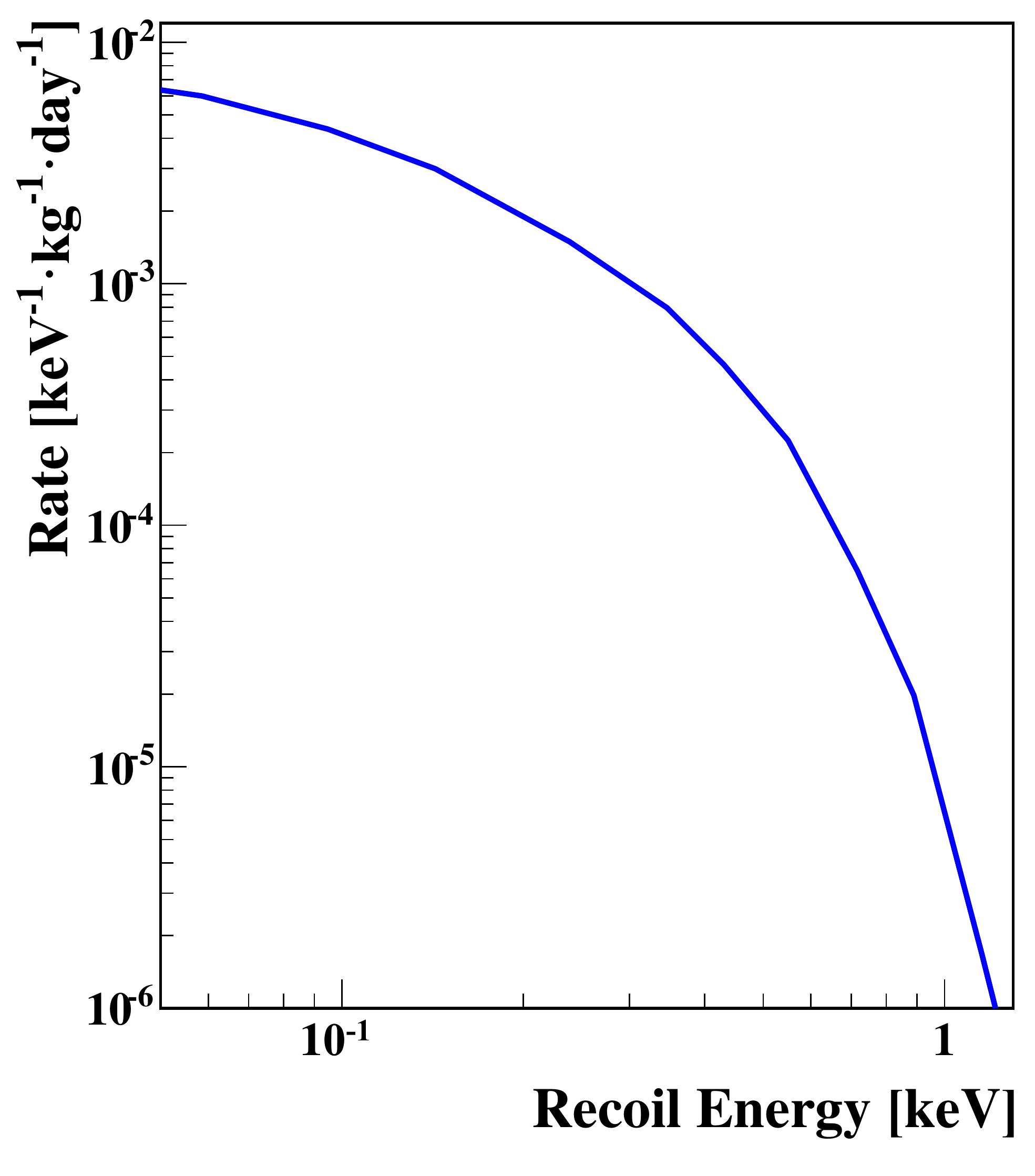}}
\centering\subfigure[]{\includegraphics[width=0.49\columnwidth]{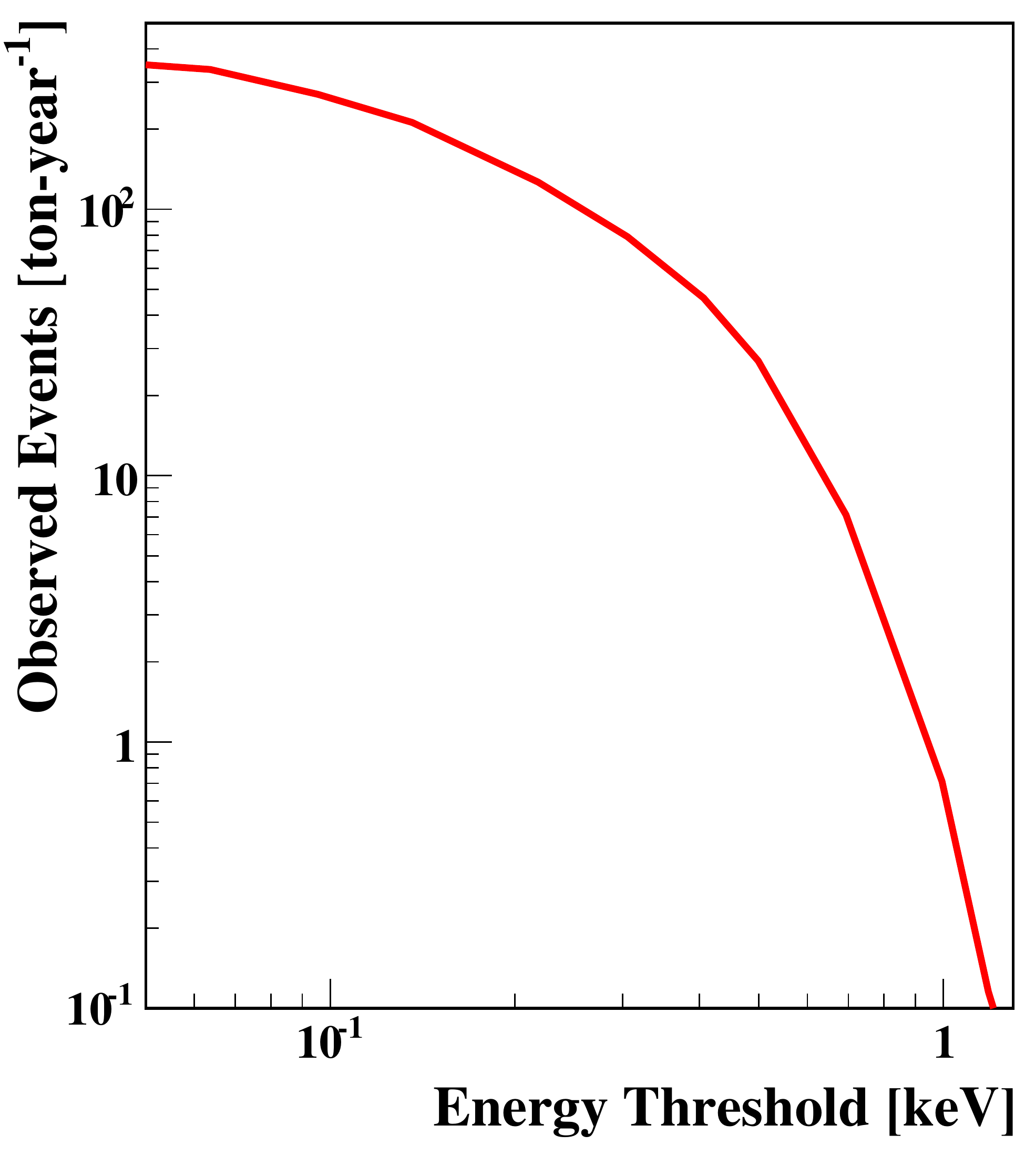}}
\caption{(a) The neutrino floor for germanium crystals as a function of recoil energy; (b) the predicted neutrino-induced events with one ton-year exposure.}
\label{fig:solar_neutrino}
\end{figure}

The first observation of coherent elastic neutrino-nucleus scattering has been reported by the COHERENT collaboration \cite{ref28} with spallation neutron source. Indistinguishable signals between the spin-independent WIMP-nucleus scattering and neutrino-nucleus coherent scattering from the solar, atmospheric, and interstellar neutrinos restrict the direct searches of WIMPs and in the meantime, provide new opportunities to study neutrino properties. Figure \ref{fig:solar_neutrino} (a) depicts the neutrino floor for germanium detectors as a function of visible ionization energy \cite{ref29}. A threshold of 200 eV with the background level of 2$\times$10$^{-3}$ cpkkd for an ionization germanium detector reaches the neutrino floor inevitably. A next generation of a tonne-scale germanium detector array has potential to detect neutrino-induced events. The predicted events with the exposure of one ton-year have been demonstrated in Figure \ref{fig:solar_neutrino} (b).

\subsubsection{$0\nu\beta\beta$ decay detection}
Several experiments aiming at $0\nu\beta\beta$ decay detection were conducted with the enriched germanium detector. GERDA achieved the background level of 1.9$^{+3.0}_{-1.4}\times 10^{-6}$ cpkkd with an energy resolution (Full Width Half Maximum, FWHM) of 3 keV at Q$_{\beta\beta}$ of 2039 keV \cite{ref3}; Majorana Demonstrator has achieved the background level of 4.9$^{+8.5}_{-3.0}\times 10^{-6}$ cpkkd with an energy resolution (FWHM) of 2.4 keV at Q$_{\beta\beta}$ \cite{ref2}. The cosmogenic background in the $0\nu\beta\beta$ decay region of interest comes from $^{60}$Co $\beta^{-}$ decay and $^{68}$Ga $\beta^{+}$ decay.

\begin{figure}[!htbp]
\centering\subfigure[]{\includegraphics[width=0.49\columnwidth]{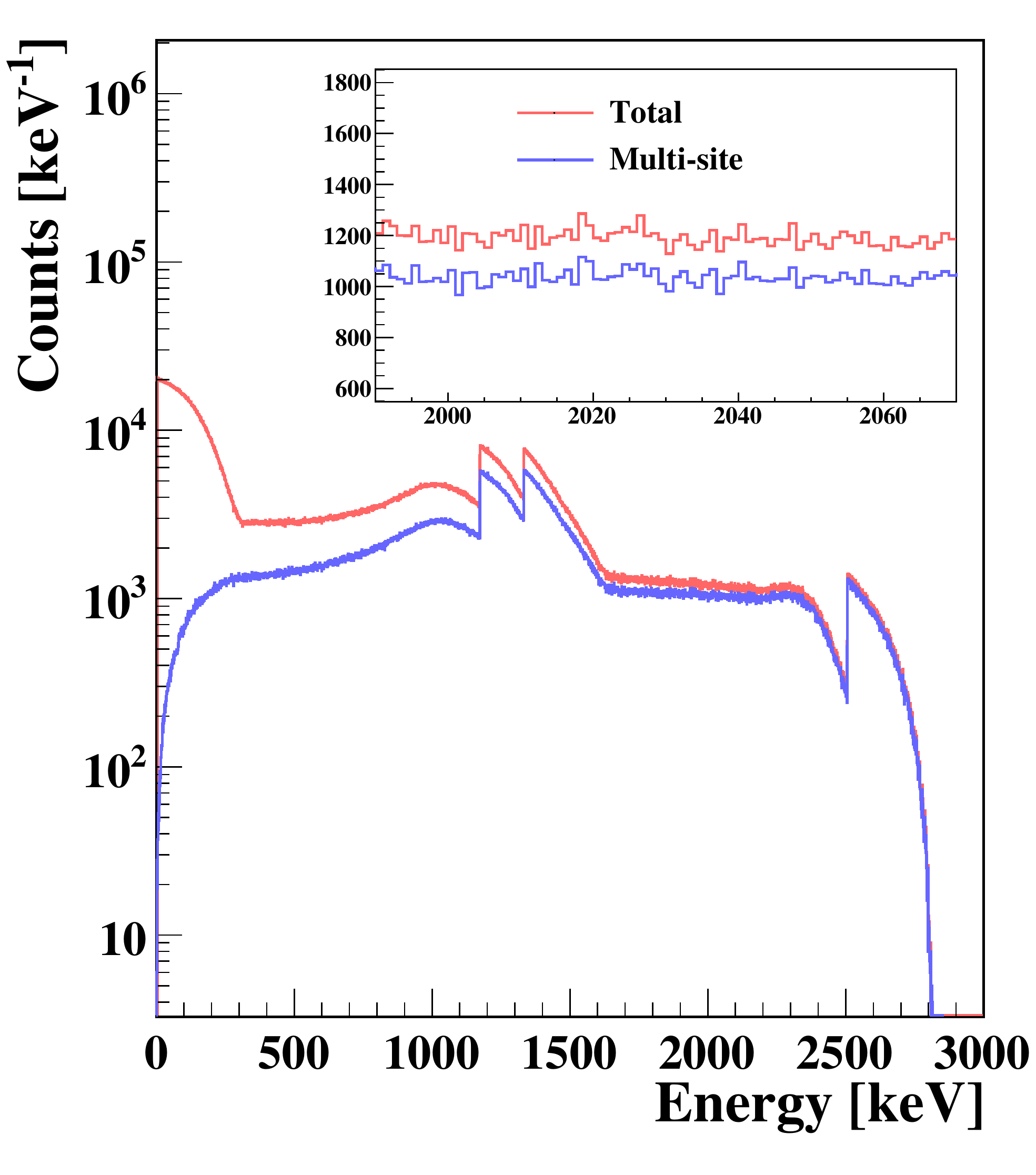}}
\centering\subfigure[]{\includegraphics[width=0.49\columnwidth]{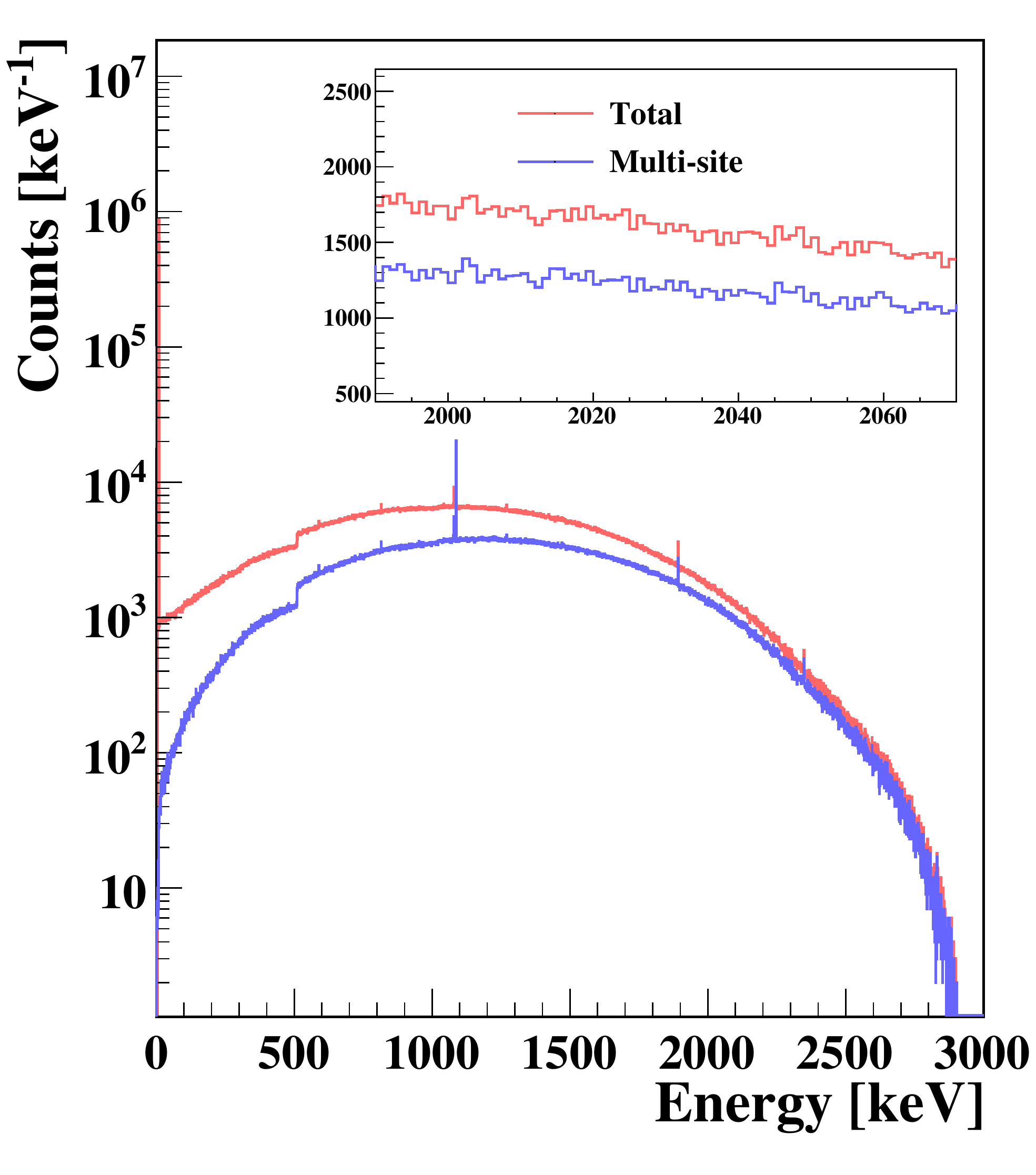}}
\caption{Spectra of $^{60}$Co (a) and $^{68}$Ga (b) in germanium crystal; the total spectra are shown as red lines, and multi-site spectra are shown as light blue lines. In the $0\nu\beta\beta$ decay region of interest, the ratios of multi-site events are more than 90\% and 70\% respectively.}
\label{fig:multi_site_ratio}
\end{figure}

The main advantage of point-contact germanium detectors is the good capability of pulse shape discrimination; that is, we can further reduce the background level by identifying and removing multi-site events \cite{ref30}. A simulation on multi-site events in germanium crystal was carried out with Geant4. The dominate decay mode of both $^{60}$Co and $^{68}$Ga are $\beta$ decay, followed by de-excitation gamma rays. The energy deposition around nuclide within and out of 1 cm was recorded, and the multi-site events defined in simulation are the events when energy is deposited in both regions. Figure~\ref{fig:multi_site_ratio} depicts the total and multi-site events spectra of $^{60}$Co and $^{68}$Ga. In the $0\nu\beta\beta$ decay region of interest, the ratios of multi-site events are more than 90\% and 70\% respectively. 

\begin{figure}[!htbp]
\centering\includegraphics[width=\columnwidth]{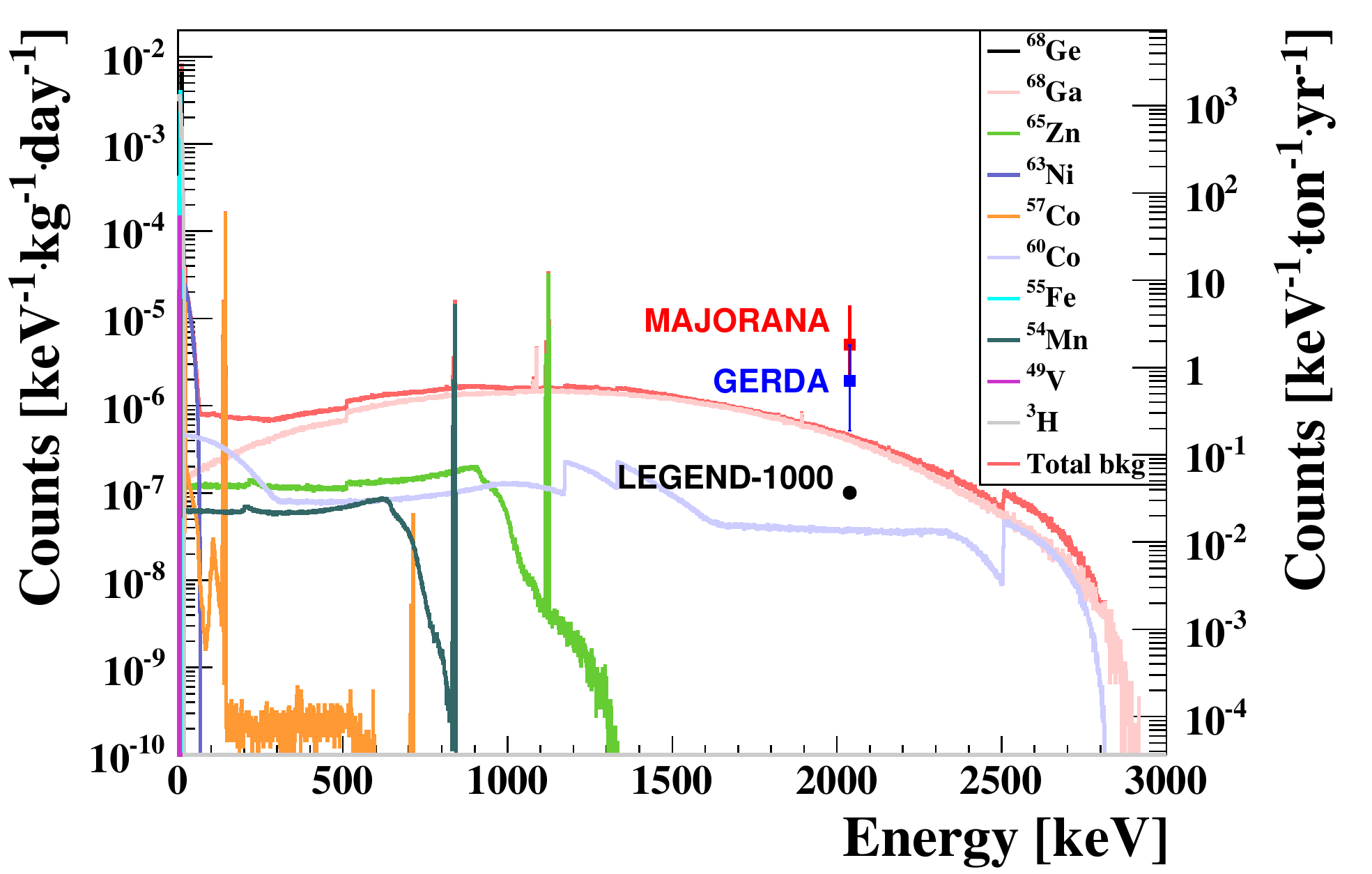}
\caption{Cosmogenic background level in $^{70}$Ge depleted germanium with 6 years cooling and the suppression index of 0.1 on $^{60}$Co and 0.3 on $^{68}$Ga in the $0\nu\beta\beta$ decay region of interest.}
\label{fig:depleted_bkg_6y}
\end{figure}

\begin{figure}[!htbp]
\centering\includegraphics[width=\columnwidth]{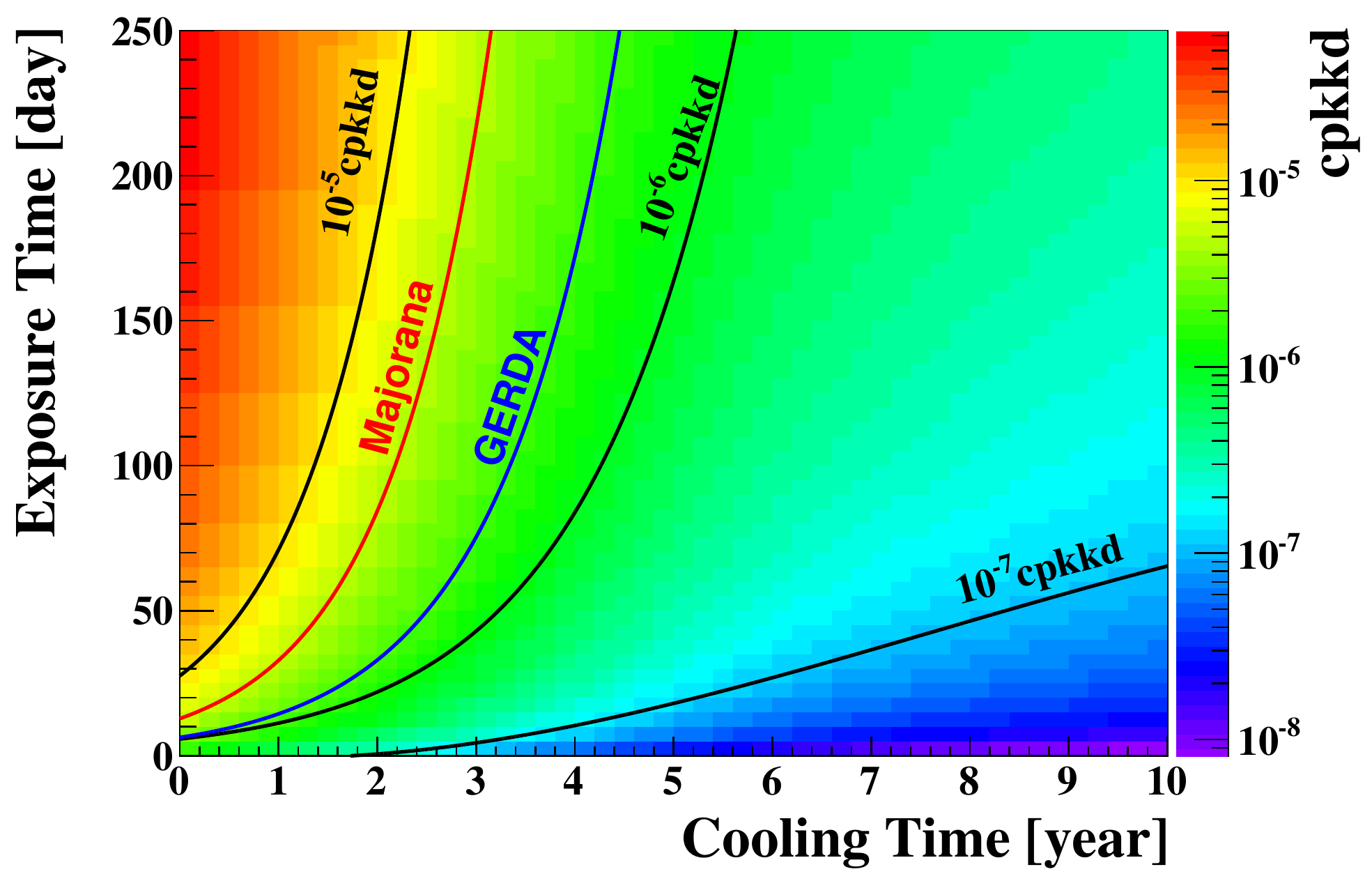}
\caption{The background level distribution in $^{70}$Ge depleted germanium detector at 2039 keV with the suppression index of 0.1 on  $^{60}$Co and 0.3 on $^{68}$Ga for $0\nu\beta\beta$ decay detection.}
\label{fig:cpkkd_in_general}
\end{figure}

Based on isotope separation and detector fabrication above ground described in section \ref{sec:depletedGe} and the assumption that multi-site events can be efficiently removed, the estimated background is about 10$^{-6}$ cpkkd with 6 years cooling, as shown in Figure~\ref{fig:depleted_bkg_6y}. However, this still cannot meet the background requirement of future tonne-scale experiment, such as LEGEND-1000 with a goal of less than 0.1 count/(FWHM$\cdot$ton$\cdot$yr) (around $10^{-7}$ cpkkd) \cite{ref31}; hence, underground crystal growth and detector fabrication are probably inevitable.

Considering underground germanium crystal growth and HPGe detector fabrication to dramatically shorten the exposure time, we evaluated the background level with an assumption of  40 days of isotope separation above ground. The background level in $^{70}$Ge depleted germanium is displayed in the ``exposure time - cooling time'' space, as shown in Figure~\ref{fig:cpkkd_in_general}. The exposure time spreads from 0 (ideal situation with underground crystal growth) to 250 days (general case for commercial detector without special treatment), and the cooling time spreads from 0 to 10 years. The suppression indices of 0.1 on $^{60}$Co and 0.3 on $^{68}$Ga were used in the region of interest. Thus, the background level can be less than 10$^{-7}$ cpkkd as cooling time extends. Several selected examples of background level are marked out in black lines. 

\section{Summary}\label{sec:5}
The cosmogenic activation in germanium was evaluated based on codes developed with Geant4 and CRY. The production rates of several long-lived radionuclides were calculated and validated by CDEX-1B experimental data. The cosmogenic background levels in both natural and $^{70}$Ge depleted germanium were predicted for future tonne-scale CDEX experiment; the levels are in the order of 10$^{-3}$ cpkkd at low-energy range and in the order of 10$^{-7}$ cpkkd around 2 MeV under the reasonable assumption of exposure time above ground and cooling time at CJPL. For direct dark matter detection, the background at low-energy range is dominated by the continuous $\beta ^-$ spectrum of $^3$H and the characteristic X-ray peaks mainly from $^{68}$Ge and $^{68}$Ga. Long cooling time at underground laboratory is inefficient to reduce the total background level, unless the complete detector fabrication is carried out underground to suppress the production of $^3$H. Thus, crystal growth and detector fabrication underground is probably the inevitable course for future tonne-scale CDEX dark matter experiment. With low background and low threshold, it also provides potential for tonne-scale germanium detector array to detect solar neutrino and $0\nu\beta\beta$ decay. For $0\nu\beta\beta$ decay detection, the continuous spectrum of $^{60}$Co $\beta^{-}$ decay and $^{68}$Ga $\beta^{+}$ decay contribute the background in the region of interest of 2039 keV. Long cooling time at underground laboratory is one way to reduce the background level; similarly, some of the detectors used in GERDA experiment have already been underground for more than 10 years. Underground crystal growth and detector fabrication is more effective and necessary to reduce the cosmogenic background with a reasonable cooling time. 




\begin{acknowledgments}
This work was supported by the National Key Research and Development Program of China (No. 2017YFA0402200 and 2017YFA0402201), the National Science Fund for Distinguished Young Scholars (No. 11725522), the National Natural Science Foundation of China (No. 11475092, 11475099, 11675088), the National Basic Research Program of China (973 Program) (No. 2010CB833006), and the Tsinghua University Initiative Scientific Research Program (No. 20151080354).
\end{acknowledgments}



\end{document}